%% file: main.tex
\pdfoutput=1
\documentclass[journal]{IEEEtran}
\usepackage{cite}
\usepackage{amsmath}
\usepackage{amsfonts}
\usepackage{graphicx}
\usepackage[caption=false, font=footnotesize]{subfig} %
\graphicspath{{figures/}}
\usepackage{booktabs} %
\usepackage[caption=false, font=footnotesize]{subfig}
\usepackage{array}
 \newcolumntype{M}[1]{>{\centering\arraybackslash}m{#1}} %
\usepackage{multirow}
\usepackage{xcolor}
\usepackage[per-mode = fraction]{siunitx} %
\usepackage{afterpage} %
\usepackage{placeins} %
\usepackage{circuitikz}

\renewcommand{\Re}{\operatorname{Re}}

\newcommand{\jw}{j\omega}
\newcommand{\w}{\omega}
\newcommand{\Y}{\mathbf{Y}}
\newcommand{\Z}{\mathbf{Z}}
\newcommand{\A}{\mathbf{A}_z}
\newcommand{\iA}{\mathbf{A}_z^{-1}}
\newcommand{\I}{\mathbf{I}}
\renewcommand{\L}{\mathbf{L}}

\newcommand{\ubL}{\underline{\mathbf{L}}}
\newcommand{\notimplies}{%
  \mathrel{{\ooalign{\hidewidth$\not\phantom{=}$\hidewidth\cr$\implies$}}}}
\newcommand{\msum}{\sum_{k=1}^n}

\newcommand{\G}{\mathbf{G}}
\renewcommand{\H}{\mathbf{H}}

\newcommand{\cnd}{\mathcal{Q}}

\newcommand{\cnds}{\mathcal{S}}

\newcommand{\us}{\overline{\sigma}}

\newtheorem{theorem}{Theorem}

\DeclareSIUnit[]\pu{p.u.} %
\newcommand{\SInp}[2]{\qty[parse-numbers = false]{#1}{#2}} %
\definecolor{clr1}{rgb}{0.11, 0.02, 0.604}
\definecolor{clr2}{rgb}{0.5, 0.055, 0.596}
\definecolor{clr3}{rgb}{0.843, 0.416, 0.102}
\makeatletter
\def\mathcolor#1#{\@mathcolor{#1}}
\def\@mathcolor#1#2#3{%
  \protect\leavevmode
  \begingroup
    \color#1{#2}#3%
  \endgroup
}
\makeatother

\title{Decentralized Impedance Specifications for Grid-Connected Converters: a Sequence-Domain Gershgorin Method, and a Frequency-Wise Framework for Combining Different Criteria}
\author{Lazar~Stojanovi\'c,~\IEEEmembership{Student Member,~IEEE,}
        Ru\v{z}ica~Cvetanovi\'c,
        Paolo~Mattavelli,~\IEEEmembership{Fellow,~IEEE}
        Simone~Buso,~\IEEEmembership{Member,~IEEE}%
\thanks{This work has been submitted to the IEEE for possible publication. Copyright may be transferred without notice, after which this version may no longer be accessible.}%
\thanks{The research leading to these results has received funding from the European Union's Horizon 2020 research and innovation programme under the Marie Sklodowska-Curie grant agreement no 101034319 and from the European Union — NextGenerationEU. This work was also partially supported by the project WIND-DIGIPOWER ``Large-scale WIND integration for the future DIGItal POWER grid using innovative power electronics control and communication-based estimations'' funded within the call ``CETPartnership Joint Call 2022''.}%
\thanks{L. Stojanovi\'c and P. Mattavelli are with the Department of Management and Engineering,
University of Padova, 36100 Vicenza, Italy (e-mail: lazar.stojanovic@phd.unipd.it; paolo.mattavelli@unipd.it).}%
\thanks{R. Cvetanovi\'c is with Innovation Center of the School of Electrical Engineering in Belgrade, 11000 Belgrade, Serbia (e-mail:
ruzica.cvetanovic@ic.etf.bg.ac.rs).}%
\thanks{S. Buso is with the Department of Information Engineering, University of Padova, 35131 Padova, Italy
(e-mail: simone.buso@dei.unipd.it).}%
}
\begin{document}

\maketitle
\input{sections/1-Introduction}

\input{sections/2-Preliminaries}
\input{sections/4-DecentralizedSpecs}
\input{sections/5-Examples}

\input{sections/6-Conclusions}

\input{sections/7-Appendix}

\bibliographystyle{IEEEtran}
\bibliography{bibtex/IEEEabrv, bibtex/SingleNodeDecAC}

\end{document}

%% file: sections/1-Introduction.tex
\begin{abstract}
    In complex grid scenarios, with many power electronics (PE) converters connected, having specifications for converters can ease the design process.
    A practical approach is to impose the specifications on the converter's small-signal impedance in the frequency domain. This is advantageous since impedance is a physical property that can be measured.
    This paper focuses on providing decentralized impedance specifications for each converter connected to an ac bus. With decentralized specifications, each converter has its own specifications, which are independent of other converters' impedance. This enables a competitive multi-vendor environment while guaranteeing system stability. Existing methods of imposing decentralized specifications, based on the small-gain, passivity and Gershgorin theorems, are formulated within a common frequency-domain framework. Building on this, two new solutions that reduce the conservativeness of the specifications are proposed. First, Gershgorin-based decentralized specifications are formulated in the modified sequence domain. Second, a framework for combining different specification types across frequency ranges is developed and mathematically proven. The methods are compared on examples covering sub-synchronous and harmonic-range interactions of grid-following converters, grid-forming converters, and a mixed grid-forming/grid-following system, using analytical models, black-box impedance measurements, and time-domain responses obtained from simulations and an experimental setup.
\end{abstract}


\newpage
\section{Introduction}

\IEEEPARstart{T}{he} global transition toward renewable energy sources has significantly increased the penetration of power electronics (PE) converters into electrical grids, introducing challenges such as converter interactions and small-signal stability issues.
Traditional stability assessment techniques---state-space pole analysis and frequency-domain control loop-gain methods---are effective for individual converter design, but become
complicated when multiple converters are interconnected, even for simple practical cases.

Impedance-based stability assessment offers an effective framework for studying
converter interactions. First applied by Middlebrook for input filter design
\cite{middlebrookInputFilterConsideration1976} and later extended to grid-connected
converters \cite{sunImpedanceBasedStabilityCriterion2011,wenInverseNyquistStability2017},
the method partitions the system into source and load subsystems, each represented
by its Thévenin or Norton equivalent. Stability is then assessed by applying the
Nyquist criterion to the source-to-load impedance ratio, known as the
\emph{minor loop gain}.

While a single-input single-output (SISO) representation is sufficient to capture the dynamics at the dc terminal of PE converters, the ac-side typically requires a multiple-input multiple-output (MIMO) representation, usually in the $dq$ frame \cite{wangHarmonicStabilityPower2019,wenImpedanceBasedAnalysisGridSynchronization2016,wenAnalysisDQSmallSignal2016} to capture all relevant dynamics. Consequently, the ac-side impedance is a transfer function matrix, which necessitates the use of the generalized Nyquist criterion.

Since applying the Nyquist criterion offers limited insight into shaping the desired individual impedances, alternative impedance-based stability specifications were first proposed by Middlebrook~\cite{middlebrookInputFilterConsideration1976} and later extended by Wildrick et al.~\cite{wildrickMethodDefiningLoad1995}.
For MIMO systems, several approaches to defining such specifications have been introduced, including methods based on matrix norms~\cite{liuInfinityNormImpedanceBasedStability2015}, on the diagonal $d$- and $q$-axis elements of the impedance matrix (applicable when the impedances are diagonally dominant)~\cite{wenACStabilityAnalysis2017}, passivity theory requiring both source and load subsystems to be passive~\cite{harneforsPassivityBasedStabilityAssessment2016}, net damping criteria requiring the combined damping of both subsystems to be positive~\cite{harneforsProofApplicationPositiveNetDamping2011,stamatiouStabilityAnalysisTwoTerminal2016,chenLimitationsUsingPassivity2024,kowalewskiPassivityBasedRobustStability2025}, and, more recently, a rotated passivity index~\cite{chenExtendedFrequencyDomainPassivity2024}.

In the case where specifications are needed for multiple converters, the decentralized impedance specifications \cite{fengImpedanceSpecificationsStable2002} were proposed. Here, the goal is to define individual conditions that each converter should satisfy independently of the others so that, once all converters are connected together, the system remains stable.
This can be of great interest when many converters are planned for connection to the grid since the specifications for each converter can be given without knowing the impedance of the others up front.

Although impedance specifications for a single converter are well established in the literature for both dc systems \cite{middlebrookInputFilterConsideration1976, wildrickMethodDefiningLoad1995} and ac systems \cite{liuInfinityNormImpedanceBasedStability2015, chenLimitationsUsingPassivity2024, chenExtendedFrequencyDomainPassivity2024}, decentralized specifications for multiple converters have predominantly been developed for dc systems---including a single dc bus \cite{fengImpedanceSpecificationsStable2002, wangDecentralizedImpedanceSpecifications2017}, series-type dc microgrids \cite{wangDecentralizedImpedanceBasedStability2019}, multi-bus dc systems \cite{zhangDecentralizedImpedanceSpecifications2024, stojanovicPowerConverterDecentralized2025}, and dc distribution systems connected to an ac grid \cite{heImpedanceSpecificationsStability2023}---and only recently extended to converters connected to a common ac bus \cite{stojanovicDecentralizedImpedanceSpecifications2024, chenAssessmentCauseLocalization2025}, which is also the focus of this paper.
For converters connected to different ac buses, decentralized conditions have been derived based on passivity of large-signal grid-forming models \cite{hePassivityDecentralizedStability2024}, on combining the small-gain and small-phase theorems \cite{huangGainPhaseDecentralized2024}, and, for multi-vendor grids, on robust stability analysis of grey-box converter models \cite{cecatiInteroperabilitySpecificationsMultiVendor2025}. Related control-theoretic conditions have also been proposed for dc microgrids \cite{laibDecentralizedStabilityConditions2023} and for generic networked systems \cite{zhaoWhenSmallGain2022, baron-pradaDecentralizedStabilityConditions2025}.
However, these conditions are formulated on control-loop transfer functions rather than on the terminal impedance \cite{cecatiInteroperabilitySpecificationsMultiVendor2025}, address large-signal rather than small-signal stability \cite{hePassivityDecentralizedStability2024}, or rely on properties, such as sectoriality, that converter impedances often do not satisfy \cite{huangGainPhaseDecentralized2024, chenAssessmentCauseLocalization2025}; imposing them directly on measured frequency data is therefore not straightforward. In contrast, for converters connected to a common ac bus, the specifications reduce to simple per-converter conditions on the measured impedance.

In our previous work \cite{stojanovicDecentralizedImpedanceSpecifications2024}, we proposed a method that formulates decentralized specifications for the $dq$-frame impedance based on the Gershgorin theorem, previously mainly employed for stability analysis \cite{liuSmallSignalStabilityCriterion2023, renStrictlySufficientStability2020, zengStabilityAssessmentMultiple2022, zengStabilityAssessmentParametric2023a, liGersgorinTheoryRobust2016}.
It is recognized in \cite{stojanovicDecentralizedImpedanceSpecifications2024} that the approach introduces a certain degree of conservativeness; however, alternative decentralized-specification methods and a systematic comparison of their conservativeness are not addressed.
More recently, \cite{chenAssessmentCauseLocalization2025} proposed decentralized specifications based on a combination of the net-damping criterion and a rotated passivity index.
Similarly to \cite{stojanovicDecentralizedImpedanceSpecifications2024}, it is demonstrated that assigning individual specifications to each converter can effectively prevent instability in multi-parallel converter systems.

Despite these advances, several gaps remain for converters connected to a common ac bus.
First, comparisons between decentralized-specification methods are limited: \cite{stojanovicDecentralizedImpedanceSpecifications2024} does not compare against alternatives, while \cite{chenAssessmentCauseLocalization2025} compares its extended-passivity condition only with conventional passivity and small-gain conditions for a grid-forming case, without considering Gershgorin-based specifications.
Second, both \cite{stojanovicDecentralizedImpedanceSpecifications2024, chenAssessmentCauseLocalization2025} primarily address low-frequency (near-synchronous) interactions, leaving the harmonic frequency range largely unexplored.
Finally, combinations of methods across frequency ranges have so far been limited to passivity- and small-gain-type conditions: \cite{chenExtendedFrequencyDomainPassivity2024, chenAssessmentCauseLocalization2025} combine the rotated passivity index at low frequencies with the net-damping criterion at high frequencies, and \cite{chenUnifiedFlexibleFrequencyDomain2025} unifies small-gain, passivity, and rotated-passivity conditions through frequency-dependent multipliers. However, Gershgorin-based specifications are not covered by this framework, and \cite{chenUnifiedFlexibleFrequencyDomain2025} is validated only on a single-converter system, without deriving per-converter specifications for multi-parallel converters.

To fill in these gaps, this paper provides the following novel contributions related to imposing decentralized impedance specifications for converters connected to the same ac bus:
\begin{itemize}
    \item Formulating the existing decentralized specification methods within a common frequency-domain framework, with easily-applicable expressions for each method, which is later used for their combination and comparison.
    \item Proposing the decentralized specifications in the modified sequence domain based on the Gershgorin theorem, extending the work in \cite{stojanovicDecentralizedImpedanceSpecifications2024}.
    \item Developing a general framework for combining different decentralized specification methods across frequency ranges, with a mathematical proof — based on the forbidden region principle — that the combination remains a sufficient condition for system stability.
    \item Comparing the conservativeness and applicability of all methods on examples covering both sub-synchronous and harmonic interactions, together with practical guidelines on which specification type is most appropriate for each frequency range.
    \item Verifying the specifications applied directly to black-box impedance measurements, without relying on analytical converter models.
\end{itemize}

%% file: sections/2-Preliminaries.tex
\section{Impedance-based Stability Assessment}

\begin{figure}
    \centering
    \includegraphics[scale=0.9]{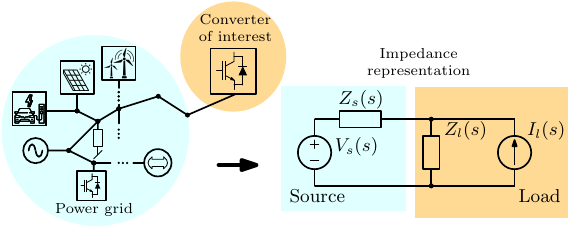}
    \caption{System partitioned to a source and load equivalent representation.}
    \label{fig3:sourceLoadRepsentation}
\end{figure}

To perform impedance-based stability assessment, the system is partitioned into source and load subsystems~\cite{middlebrookInputFilterConsideration1976}, each represented by its Th\'{e}venin or Norton equivalent with terminated small-signal impedances $Z_s(s)$ and $Z_l(s)$, as shown in Fig.~\ref{fig3:sourceLoadRepsentation}. One of the key advantages of this approach is that impedances are physical quantities that can be measured directly.

Throughout this paper, dc-system impedances are represented as SISO transfer functions, while ac-system impedances are modeled in the $dq$ frame as transfer function matrices \cite{wenImpedanceBasedAnalysisGridSynchronization2016, wangHarmonicStabilityPower2019}, denoted in bold (e.g., $\mathbf{Z}_g$), with elements:
\begin{equation}
    \Z_g(s) = \begin{bmatrix}
        Z_g^{dd}(s) & Z_g^{dq}(s) \\
        Z_g^{qd}(s) & Z_g^{qq}(s)
    \end{bmatrix}.
\end{equation}

\subsection{Applying the Nyquist criterion}
For a dc system, stability is assessed by applying the Nyquist criterion to the minor loop gain $L(s) = Z_s(s)/Z_l(s)$. Following \cite{sunImpedanceBasedStabilityCriterion2011, wenInverseNyquistStability2017}, the source is assumed stable under ideal load
(constant current) and the load stable under ideal source (constant voltage), so that $L(s)$ has no RHP poles. The system is then stable if and only if the Nyquist plot of $L(\jw)$ does not encircle $(-1,0)$. This assumption of no RHP poles in the minor loop gain is maintained throughout the paper.

For ac systems, the generalized Nyquist criterion is applied to the matrix minor loop gain $\mathbf{L}(s)$, under the same assumption of no RHP poles. The system is stable if the eigenloci of $\mathbf{L}(\jw) = \mathbf{Z}_l^{-1}(\jw)\mathbf{Z}_s(\jw)$ do not encircle $(-1,0)$.

\subsection{Forbidden region principle}
Instead of judging the stability by counting the encirclements of the $(-1,0)$ point, an alternative way is to guarantee the stability by using the forbidden region principle. The forbidden region is defined as the area in the complex plane where the frequency response of the minor loop gain $L(\jw)$ (or the eigenloci of $\L(\jw)$ in MIMO case) is not allowed to enter. The forbidden region is constructed such that staying outside of it means that no encirclement of the $(-1,0)$ point is possible, which ensures stability under no RHP poles assumption.

%% file: sections/4-DecentralizedSpecs.tex
\section{Decentralized Specifications for Converters Connected to a Single Bus}

Stability analysis of the system gives a conclusion about the stability of the interconnected system, but it offers limited insight into the design of individual converters. This was addressed by introducing impedance specifications as guidelines that designers should follow \cite{middlebrookInputFilterConsideration1976, wildrickMethodDefiningLoad1995, liuInfinityNormImpedanceBasedStability2015, harneforsPassivityBasedStabilityAssessment2016, chenExtendedFrequencyDomainPassivity2024}.

However, impedance specifications presented in \cite{middlebrookInputFilterConsideration1976, wildrickMethodDefiningLoad1995, harneforsProofApplicationPositiveNetDamping2011, chenLimitationsUsingPassivity2024}, like stability analysis, require complete knowledge of the system, and specifications can be given for only one converter while keeping the other ones fixed. This is impractical, as the specifications given to one converter depend on the rest of the grid, and changing it will result in a change of the specifications.

\begin{figure}
\centering
\includegraphics[scale=0.85, page=2]{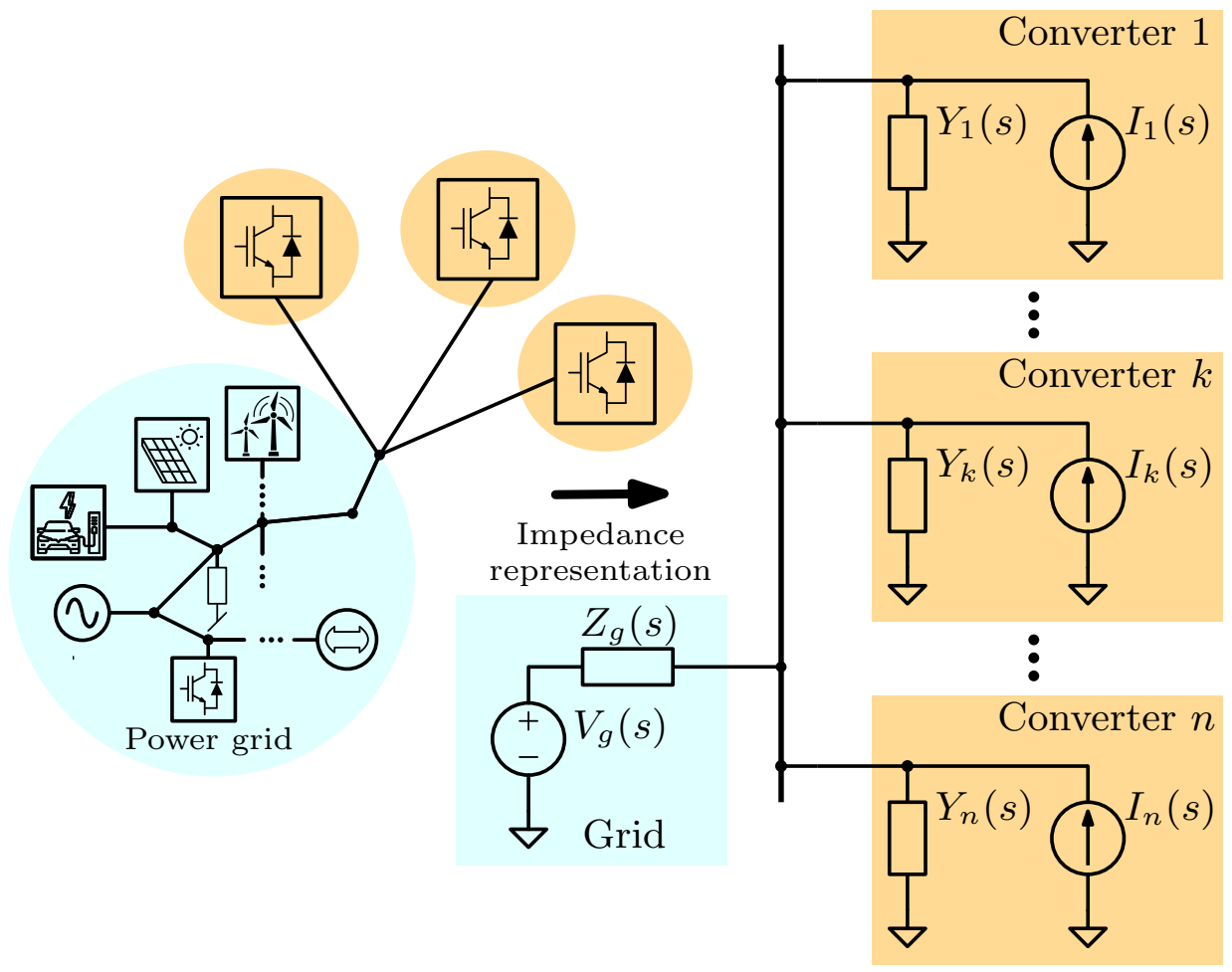}
\caption{Different converters connected to the single bus - small-signal impedance representation.}
\label{fig4:convertersConnectedToBus}
\end{figure}

This is especially problematic in a multi-vendor environment, where the converters need to be designed independently, but in such a way that the system stability is not compromised \cite{cecatiInteroperabilitySpecificationsMultiVendor2025, stojanovicDecentralizedImpedanceSpecifications2024, stojanovicPowerConverterDecentralized2025, huangGainPhaseDecentralized2024, hePassivityDecentralizedStability2024}.
This is a motivation for developing decentralized impedance specifications. In this section we will focus on multiple converters connected in parallel to a single bus, as in typical grid-tied applications (Fig.~\ref{fig4:convertersConnectedToBus}).\footnote{In recent studies, the decentralized specifications for converters connected to different buses are also considered for dc \cite{stojanovicPowerConverterDecentralized2025} and ac systems \cite{huangGainPhaseDecentralized2024, hePassivityDecentralizedStability2024}. Also, a different approach that observes the problem from a robust stability perspective was recently presented in \cite{cecatiInteroperabilitySpecificationsMultiVendor2025}.}
To emphasize the fact that we are focusing on grid-connected converters in this paper, the \emph{source} impedance is considered to be known and referred to as \emph{grid} impedance $\Z_g$, while the \emph{load} admittance is considered to be the admittance of the \emph{converter} $\Y_k$, as shown in Fig.~\ref{fig4:convertersConnectedToBus}.

The main idea of decentralized specifications is to impose individual conditions on each converter admittance, dependent only on $\Y_k(s)$ and the grid impedance $\Z_g(s)$, that ensure system stability.

To formulate the decentralized specifications mathematically, we start from the sufficient stability condition $\cnds$ of the system.
Then, the decentralized specifications are expressed as $n$ individual conditions, where each condition $\cnd_k$ corresponds to a converter $k$. The satisfaction of all individual conditions $\cnd_k$ guarantees the stability of the system:
\begin{equation}
\label{eq4:decentralizedSpecsCnds}
\cnd_1 \land \cnd_2 \land \cdots \land \cnd_k \land \cdots \land \cnd_n \to \cnds \to \text{Stability}.
\end{equation}
The condition for an individual converter $\cnd_k$ is dependent only on the admittance of that converter $\Y_k(s)$ and the grid impedance $\Z_g(s)$, and independent of the admittance of other converters.
Based on the condition's $\cnd_k$ structure, we differentiate between two types of decentralized specifications:
\begin{itemize}
\item \emph{Explicit decentralized specifications}, where specifications are given directly to converter's admittance $\Y_k(s)$.
\begin{equation}
\label{eq4:explicitDecentralizedSpecs}
\cnd_k:\ f(\Y_k) < g(M_k,\Z_g).
\end{equation}
\item \emph{Implicit decentralized specifications}, where specifications are not given directly to the converter's admittance, but nevertheless they depend only on one converter admittance at a time.
\begin{equation}
\label{eq4:implicitDecentralizedSpecs}
\cnd_k:\ f(\Y_k,\Z_g) < g(M_k).
\end{equation}
\end{itemize}
The functions $f$ and $g$, as well as parameter $M_k$, hereafter referred to as scaling factors, must be chosen in a way that guarantees \eqref{eq4:decentralizedSpecsCnds} is satisfied.
The scaling factor $M_k$ can be chosen based on the properties of the converter that are known upfront. A reasonable choice, proposed in \cite{fengImpedanceSpecificationsStable2002}, is to scale the converters based on their rated power $S_k$, $M_k = \frac{S_k}{\sum_{k=1}^n S_k}$, such that converters rated for more power are allocated a larger ``budget'', and this scaling will be used throughout the paper.

The explicit decentralized specifications give more insight into the design of the converter, however they tend to be more conservative.
On the other hand, the implicit decentralized specifications are not direct specifications on the converter's admittance. Nevertheless, they are a valuable tool for systems with a large number of converters, especially when those converters come from multiple vendors. By providing an independent condition for each converter (whether explicit or implicit) decentralized specifications foster a competitive multi-vendor environment for converter designers, while guaranteeing system stability.

\subsection{Decentralized specifications based on Middlebrook criterion}

By utilizing the small-gain theorem, the sufficient stability condition $\cnds$ for the whole system can be expressed as:
\begin{equation}
\cnds:\ \forall\w, \ \us[\Y_{eq}(\jw)] < \us[\Z_g(\jw)]^{-1}.
\label{eq4:-S-smallGain}
\end{equation}
By using the fact that
\begin{equation}
\us(\Y_{eq}(\jw)) = \us\left[\sum_{k=1}^n \Y_k(\jw)\right] \leq \sum_{k=1}^n \us[\Y_k(\jw)],
\end{equation}
decentralized specifications can be given in the form of:
\begin{equation}
\label{eq4:dec-spec-middlebrook-matrix}
\cnd_k:\ \forall\w, \ \us[\Y_k(\jw)] < \frac{M_k}{\us[\Z_g(\jw)]},
\end{equation}
where $M_k$ is the scaling factor, unique for each converter, such that $\sum_{k=1}^n M_k \leq 1$, which is an assumed property of $M_k$ throughout the paper.

\subsection{Decentralized specifications based on Passivity criterion}

A stable transfer function matrix $\G(s)$ is passive if its Hermitian part is positive definite at all frequencies, $\G(\jw)+\G^H(\jw)>0$, where $(\cdot)^H$ is the conjugate transpose \cite{desoerFeedbackSystemsInputoutput2009}. The degree of passivity at frequency $\w$ is quantified by the input feedforward passivity index \cite{chenLimitationsUsingPassivity2024}:
\begin{equation}
\label{eq4:passivityIndex}
\nu_F\left[\G,\w\right]=\frac{1}{2} \min \lambda \left[ \G(\jw) + \G^H(\jw) \right].
\end{equation}
A feedback connection of two passive subsystems is stable \cite{desoerFeedbackSystemsInputoutput2009}. The net damping criterion \cite{harneforsProofApplicationPositiveNetDamping2011}, extended to transfer function matrices in \cite{chenLimitationsUsingPassivity2024}, relaxes this requirement: a passivity deficit of the converters is allowed as long as it is compensated by the passivity surplus of the grid. With $\Y_g = \Z_g^{-1}$, the sufficient stability condition $\cnds$ for the whole system is:
\begin{equation}
\cnds:\ \forall\w, \ \nu_F\left[\Y_{eq},\w\right] > -\nu_F\left[\Y_g,\w\right] \ \land \ \nu_F\left[\Y_g,\w\right] > 0.
\end{equation}
To split this condition into individual conditions for each converter, $\cnd_k$ we use the property:
\begin{equation}
\nu_F\left[\Y_{eq},\w\right] = \nu_F\left[\sum_{k=1}^n \Y_k,\w\right] \geq \sum_{k=1}^n \nu_F\left[\Y_k,\w\right],
\end{equation}
that follows from Weyl's inequality for Hermitian matrices, $\lambda_{\min}(\mathbf{A}+\mathbf{B})\geq\lambda_{\min}(\mathbf{A})+\lambda_{\min}(\mathbf{B})$ \cite{hornMatrixAnalysis1985}.
Then, assuming that the grid is passive, the decentralized specifications based on the passivity criterion can be expressed as:
\begin{equation}
\label{eq4:dec-spec-passivity-matrix}
\cnd_k:\ \forall\w, \ \nu_F\left[\Y_k,\w\right] > -M_k\nu_F\left[\Y_g,\w\right].
\end{equation}
The special case $M_k = 0$ corresponds to requiring every converter to be passive, $\nu_F\left[\Y_k,\w\right] > 0$.

\subsection{Decentralized specifications based on Rotated passivity}

A further extension is the rotated passivity approach~\cite{chenExtendedFrequencyDomainPassivity2024}, which applies a rotation matrix $\mathbf{R}$ to the admittances before evaluating the passivity index.
While $\mathbf{R}$ can in principle be chosen arbitrarily, the choice $\mathbf{R} = \begin{bmatrix} 0 & -1 \\ 1 & 0 \end{bmatrix}$ is used in~\cite{chenExtendedFrequencyDomainPassivity2024} because it maps the inductive grid impedance to a resistive form in the rotated representation, making the criterion applicable in the low-frequency range. This $\mathbf{R}$ is adopted throughout the rest of this paper.
Analogously to \eqref{eq4:dec-spec-passivity-matrix}, the decentralized specifications based on the rotated passivity approach (also studied in \cite{chenAssessmentCauseLocalization2025}) can be expressed as:
\begin{equation}
\label{eq4:dec-spec-rotated-passivity-matrix}
\cnd_k:\ \forall\w, \ \nu_F\left[\mathbf{R}\Y_k,\w\right] > -M_k\nu_F\left[\mathbf{R}\Y_g,\w\right] \ \land \ \nu_F\left[\mathbf{R}\Y_g,\w\right] > 0.
\end{equation}
Note that a passive grid does not imply that $\nu_F\left[\mathbf{R}\Y_g,\w\right] > 0$, which should be checked separately when using the rotated passivity approach.

\subsection{Decentralized specifications based on minor loop gain}
\label{sec4:minorLoopGain}

The above-mentioned solutions for imposing decentralized specifications are in the explicit form, as in \eqref{eq4:explicitDecentralizedSpecs}.
As mentioned before, these specifications give more information to converter designers, but they tend to be more conservative.
To overcome this problem the decentralized specifications given directly to the minor loop are proposed for a \emph{dc} bus in \cite{fengImpedanceSpecificationsStable2002}.
Instead of giving the specification to each converter admittance, the specifications are given for the minor loop gain, defined for each converter $k$ as:
\begin{equation}
L_k(s) = Y_k(s)Z_g(s).
\end{equation}
For the dc case, the \emph{sufficient} stability condition is:
\begin{equation}
\cnds:\ \forall\w, \ \Re\left\{ L(\jw) \right\} = \Re\left\{ \sum_{k=1}^n L_k(\jw) \right\} > -1.
\end{equation}
Then, by using $\Re\{z_1 + z_2\} = \Re\{z_1\} + \Re\{z_2\}$, the decentralized specifications can be expressed as:
\begin{equation}
\cnd_k:\ \forall\w, \ \Re\left\{ Y_k(\jw) Z_g(\jw) \right\} > -M_k.
\end{equation}
The form of these specifications is implicit, as in \eqref{eq4:implicitDecentralizedSpecs}.

\subsubsection{Decentralized specifications based on Gershgorin theorem}
\hfil\\
Extending this idea to the ac case, where the total loop gain $\L(s) = \sum_k \L_k(s)$ is a transfer function matrix with $\L_k = \Y_k \Z_g$, is not straightforward.
The challenge is that, for non-Hermitian matrices, the eigenvalues of
the sum are not directly related to those of the individual terms:
\begin{equation}
\begin{split}
\min \Re\left\{ \lambda \left[\sum_{k=1}^n \L_k(\jw)\right] \right\} &> -1 \notimplies\\
\notimplies \sum_{k=1}^n \min \Re\left\{ \lambda \left[\L_k(\jw)\right] \right\} &> -1.
\end{split}
\end{equation}

To deal with this problem, the solution initially proposed in \cite{stojanovicDecentralizedImpedanceSpecifications2024}, and further explored in this article, uses the Gershgorin theorem. This theorem involves estimating the position of eigenvalues based on individual matrix elements, which enables the definition of decentralized specifications.

\begin{figure}[!t]
\centering
\includegraphics[scale = 0.8]{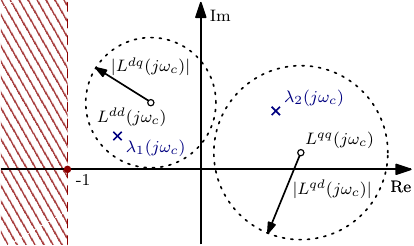}
\caption{Graphical illustration of Gershgorin theorem at frequency $\omega_c$. According to the Gershgorin theorem, the eigenvalues ($\lambda_1, \lambda_2$) of $\L(\jw_c)$ must lie within the two circles \cite{stojanovicDecentralizedImpedanceSpecifications2024}.}
\label{fig:AC-gersh}
\end{figure}

According to \cite{stojanovicDecentralizedImpedanceSpecifications2024}, by applying the Gershgorin theorem, the sufficient stability conditions are:%
\begin{equation}
\begin{split}
\cnds:\ \forall\w,\ &\Re \big\{L^{dd}(\jw) \big\} - \big|L^{dq}(\jw)\big| > -1 \ \land \\
&\Re \big\{L^{qq}(\jw) \big\} - \big|L^{qd}(\jw)\big| > -1.
\end{split}
\label{eq:GE-crit}
\end{equation}
The graphical illustration of the application of the Gershgorin theorem, for imposing specifications at frequency $\omega_c$ is given in Fig. \ref{fig:AC-gersh}.

For decentralized assessment, the loop gain is partitioned as:
\begin{equation}
\L(\jw) = \msum \ubL_k(\jw) = \begin{bmatrix}
\sum L^{dd}_k(\jw) & \sum L^{dq}_k(\jw)\\
\sum L^{qd}_k(\jw) & \sum L^{qq}_k(\jw)\\
\end{bmatrix}.
\label{eq:loop-gain}
\end{equation}
Then according to \cite{stojanovicDecentralizedImpedanceSpecifications2024} the decentralized specifications are:
\begin{equation}
\begin{split}
\cnd_k:\ \forall\w, \
&\Re \big\{L^{dd}_k(\jw) \big\} - \big|L^{dq}_k(\jw)\big| > -M_k^d \ \land \\
&\Re \big\{L^{qq}_k(\jw) \big\} - \big|L^{qd}_k(\jw)\big| > -M_k^q \, ,
\end{split}
\label{eq:Dec-spec}
\end{equation}
where the scaling factors $M_k^d$ and $M_k^q$ must satisfy:
\begin{equation}
\msum M_k^d \leq 1 \ \land \ \msum M_k^q \leq 1
\label{eq:M-cond}.
\end{equation}
The proposed criterion enables the selection of the scaling factors for the $d$ and $q$ channels independently, as long as \eqref{eq:M-cond} is satisfied.

\subsubsection{Reducing the conservativeness of the Gershgorin based decentralized specifications}
\hfil\\
The Gershgorin theorem enables us to define well-posed decentralized specifications, but, at the same time, increases the conservativeness of the assessment. This is due to the fact that we are defining the circle where eigenvalues can be located, and considering the worst-case scenario \cite{stojanovicDecentralizedImpedanceSpecifications2024,liGersgorinTheoryRobust2016}. The conservativeness is especially pronounced when the loop gain has a strong coupling between $d$ and $q$ channels (non-diagonally dominant)~\cite{stojanovicDecentralizedImpedanceSpecifications2024,liGersgorinTheoryRobust2016}.

A possible way to reduce conservativeness is to apply a similarity transformation $\A$ (with inverse $\iA$) to the loop gain~\cite{renStabilityAssessmentGridConnected2020}:
\begin{equation}
\L^t(\jw) = \A \L(\jw) \iA,
\label{eq:Lt_def}
\end{equation}
which preserves eigenvalues \cite{ryggModifiedSequenceDomainImpedance2016,zhouNovelForbiddenRegionBasedStability2019}.

In the context of decentralized specifications, according to \eqref{eq:loop-gain}, the transformed loop gain is given by:
\begin{equation}
\L^t(\jw) = \msum \A \L_k(\jw) \iA.
\end{equation}
This means that if the same arbitrary transformation $\A$ is applied to each converter's minor loop gain $\L_k(\jw)$, applying \eqref{eq:Dec-spec} to the newly transformed quantities still guarantees stability. The transformation of the system ensures that eigenvalues are conserved, but at the same time, it changes the evaluation based on Gershgorin theorem since individual elements of $\L(\jw)$ are now changed.

The main challenge is selecting the right transformation that minimizes the conservativeness of the Gershgorin theorem assessment. This is not an easy task, since the transformation must be predetermined and the same for all converters.

A special case of transformation called the modified sequence domain is presented in \cite{ryggModifiedSequenceDomainImpedance2016}. This transformation is defined by:
\begin{equation}
\A = \frac{1}{\sqrt{2}} \begin{bmatrix}
1 & j \\ 1& -j
\end{bmatrix}.
\end{equation}
One of the notable properties of this transformation is that it maps the symmetric
$dq$-frame impedance of balanced three-phase elements to diagonal form in the
modified sequence domain. For example, a three-phase inductance $L_i$ whose $dq$
impedance is
\begin{equation}
\Z_L^{dq}(s) = \begin{bmatrix} sL_i & -\omega_s L_i \\ \omega_s L_i & sL_i \end{bmatrix},
\end{equation}
becomes diagonal in the modified sequence domain:
\begin{equation}
\Z_L^{pn}(s) = \A \Z_L^{dq}(s) \A^{-1} = \begin{bmatrix} (s+j\omega_s)L_i & 0\\ 0 & (s-j\omega_s)L_i \end{bmatrix}.
\end{equation}
This diagonality reduces the off-diagonal (coupling) terms of the loop gain, which
are directly responsible for the conservativeness of the Gershgorin assessment.

In this paper, to distinguish between the analyses in the $dq$ domain and the modified sequence domain, the notation for quantities in the modified sequence domain will use the superscript $pn$ instead of $dq$. For example, loop gain will be written as:
\begin{equation}
\L^{pn}(\jw) = \A \L^{dq}(\jw) \A^{-1} =
\begin{bmatrix}
L^{pp}(\jw) & L^{pn}(\jw) \\
L^{np}(\jw) & L^{nn}(\jw)
\end{bmatrix}.
\end{equation}
From here, the dual expression to \eqref{eq:Dec-spec} for decentralized specifications in the modified sequence domain is obtained:
\begin{equation}
\begin{split}
\cnd_k:\ \forall\w,\ &\Re \Big\{L^{pp}_k(\jw) \Big\} - \big|L^{pn}_k(\jw)\big| > -M_k^p \ \land \\
&\Re \Big\{L^{nn}_k(\jw) \Big\} - \big|L^{np}_k(\jw)\big| > -M_k^n \, ,
\end{split}
\label{eq:Dec-spec_pn}
\end{equation}
where the $M_k^p$ and $M_k^n$ can be selected independently in the same way as $M_k^d$ and $M_k^q$ with conditions:
$
\msum M_k^p \leq 1 \ \land \ \msum M_k^n \leq 1.
\label{eq:M-cond_pn}
$

For a system consisting only of symmetric subsystems, such as RLC branches, each local loop gain in the modified sequence domain is diagonal, i.e., $L^{pn}_k = L^{np}_k = 0$. The Gershgorin circles then have zero radius, the eigenvalue locations are known exactly, and \eqref{eq:Dec-spec_pn} is never more conservative than \eqref{eq:Dec-spec}. Converters are generally not symmetric, particularly at low frequencies; nevertheless, the case studies in Section~\ref{sec:caseStudies} show that the conservativeness can still be reduced by using the modified sequence domain. In the high-frequency range, the influence of asymmetric control loops (e.g., PLL and active and reactive power control) diminishes, so the modified sequence domain is generally preferred there.

\subsection{Combination of different decentralized specifications}

Although satisfying one type of decentralized specifications is sufficient for the stability of the system, it is not necessary, and can lead to limited applicability of the specifications. For example, it is well known that the passivity-based specifications are almost impossible to satisfy for low frequencies in realistic scenarios \cite{chenExtendedFrequencyDomainPassivity2024}. Because of this, using different types of specifications for different frequency ranges is of interest.

To justify combining different specification types across frequency ranges, we must verify that the combination still constitutes a sufficient condition for stability.

To formalize this, instead of defining the conditions as $\cnd_k$ for the whole frequency range, we need them to be defined for each frequency $\w$, as $\cnd_k(\w)$. The type of specification is denoted by the superscript $\theta \in \Theta = \{\mathrm{M}, \mathrm{P}, \mathrm{RP}, dq, pn\}$, so that $\cnd_k^{\mathrm{M}}(\w)$, $\cnd_k^{\mathrm{P}}(\w)$, $\cnd_k^{\mathrm{RP}}(\w)$, $\cnd_k^{dq}(\w)$, and $\cnd_k^{pn}(\w)$ are the conditions \eqref{eq4:dec-spec-middlebrook-matrix}, \eqref{eq4:dec-spec-passivity-matrix}, \eqref{eq4:dec-spec-rotated-passivity-matrix}, \eqref{eq:Dec-spec}, and \eqref{eq:Dec-spec_pn}, respectively, evaluated at the single frequency $\w$.
For example, $\cnd_k^{\mathrm{P}}(\w)$ is defined on frequency $\w$, based on \eqref{eq4:dec-spec-passivity-matrix}, as:
\begin{equation}
\cnd_k^{\mathrm{P}}(\w):\ \nu_F\left[\Y_k,\w\right] > -M_k\nu_F\left[\Y_g,\w\right].
\end{equation}
The other frequency dependent conditions are defined in a similar way.

\begin{theorem}
\label{thm:combination}
Assume that the minor loop gain $\L(s) = \Y_{eq}(s)\Z_g(s)$ has no RHP poles and that $\msum M_k \leq 1$ holds for every type of specifications. If at each frequency at least one specification type is satisfied by all converters, i.e.,
\begin{equation}
\label{eq:combination}
\forall \w,\ \exists \theta \in \Theta: \ \cnd_1^{\theta}(\w) \land \cnd_2^{\theta}(\w) \land \cdots \land \cnd_n^{\theta}(\w),
\end{equation}
then the system is stable.
\end{theorem}

\begin{IEEEproof}
The proof is done by identifying a common forbidden region shared by all specification types, as illustrated in Fig.~\ref{fig:combined_forbidden_regions}.

\begin{figure}
\centering
\includegraphics[scale=0.8]{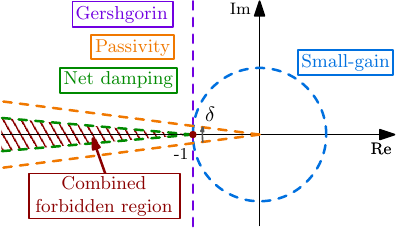}
\caption{Graphical illustration of the forbidden region principle for different types of specifications.}
\label{fig:combined_forbidden_regions}
\end{figure}

For small-gain specifications it is well known that fulfilling \eqref{eq4:-S-smallGain} keeps the eigenvalues of $\L(\jw)$ inside the unit circle, as illustrated in Fig.~\ref{fig:combined_forbidden_regions}.

Gershgorin based specifications are designed to keep the real part of eigenvalues of $\L(\jw)$ greater than $-1$, as discussed in Fig.~\ref{fig:AC-gersh}.

For passivity-based specifications, the corresponding forbidden region is less direct. It can be obtained via the sectorial decomposition and phase properties of complex matrices presented in \cite{wangPhasesComplexMatrix2020}. For our case the sectorial decomposition of the $\Z_g(\jw)$ gives the two angles $\phi_1(\jw)$ and $\phi_2(\jw)$, and they are ordered such that $\overline{\phi}(\jw) = \phi_1(\jw) \geq \phi_2(\jw) = \underline{\phi}(\jw)$ \cite{wangPhasesComplexMatrix2020}. These angles form a sector in the complex plane $(\underline{\phi}(\jw), \overline{\phi}(\jw))$, and all eigenloci of $\Z_g(\jw)$ must lie within this sector. The sectorial decomposition exists when the numerical range of the matrix does not contain the origin, which is always the case for strictly passive matrices \cite{wangPhasesComplexMatrix2020}. Or in other words, the phase of the eigenvalues of $\Z_g$ must be in between $\underline{\phi}(\jw)$ and $\overline{\phi}(\jw)$: $\arg\{\lambda[\Z_g(\jw)]\} \in (\underline{\phi}(\jw), \overline{\phi}(\jw))$ \cite{wangPhasesComplexMatrix2020}. For our proof we will use the following properties:
\begin{itemize}
\item If we have two complex matrices $\Z_g$, with sector $(\phi_2, \phi_1)$ and $\Y_{eq}$, with sector $(\psi_2, \psi_1)$, then the phase of the eigenvalues of $\L = \Y_{eq}\Z_g$ must be located in the sector $(\phi_2 + \psi_2, \phi_1 + \psi_1)$ \cite{wangPhasesComplexMatrix2020}.
\item If matrix $\Z_g$ is passive ($\Z_g + \Z_g^H> 0 \Leftrightarrow \nu_F[\Z_g]>0$) then the sector of $\Z_g$ must be fully contained in ($-\pi/2, \pi/2$) \cite{wangPhasesComplexMatrix2020}.
\end{itemize}
Note also that passivity is preserved under inversion: if $\mathbf{A}+\mathbf{A}^H>0$, then $\mathbf{A}^{-1}+\mathbf{A}^{-H} = \mathbf{A}^{-1}\left(\mathbf{A}+\mathbf{A}^H\right)\mathbf{A}^{-H}>0$. Hence, $\nu_F[\Y_g,\w]>0$ implies $\nu_F[\Z_g,\w]>0$.

We first consider the special case $M_k=0$ of \eqref{eq4:dec-spec-passivity-matrix}, where all converters, and therefore $\Y_{eq}$, are passive. If we assume that both $\Z_g(\jw)$ and $\Y_{eq}(\jw)$ are passive, then the eigenloci of $\L(\jw) = \Y_{eq}(\jw)\Z_g(\jw)$ must be located in the sector $(-\pi/2-\pi/2, \pi/2+\pi/2) = (-\pi, \pi)$. In other words, the eigenloci of $\L(\jw)$ can't enter the wedge $(\pi-\delta/2, -\pi+\delta/2)$ with infinitely small angle $\delta > 0$ around the negative real axis, as illustrated in Fig.~\ref{fig:combined_forbidden_regions}.

For the general case $M_k>0$ of \eqref{eq4:dec-spec-passivity-matrix}, i.e., net damping specifications, the situation is different, since $\nu_F[\Y_{eq},\omega]>0$ is not guaranteed. Then we define:
\begin{equation}
\H(\jw) = \I + \L(\jw) = [\Y_g(\jw) + \Y_{eq}(\jw)]\Z_g(\jw).
\end{equation}
Since $\nu_F[\Y_{eq},\, \w]+ \nu_F[\Y_{g},\, \w]>0$ then it is also true that $\nu_F[\Y_g + \Y_{eq},\, \w]>0$. And since $\Z_g(\jw)$ is passive, the eigenvalues of $\H(\jw)$ can't be located in the wedge $(\pi-\delta/2, -\pi+\delta/2)$. Because eigenvalues of $\L(\jw)$ are just eigenvalues of $\H(\jw)$ shifted by $-1$, the eigenvalues of $\L(\jw)$ can't be located in the wedge $(\pi-\delta/2, -\pi+\delta/2)$ around the point $-1$ on the real axis, as illustrated in Fig.~\ref{fig:combined_forbidden_regions}.

Rotated passivity specifications can be analyzed in the same way. For invertible matrix $\mathbf{R}$, we can transform the loop gain without changing the eigenvalues, as in \eqref{eq:Lt_def}:
\begin{equation}
\L = \Y_{eq}\Z_g \to \L^t = \mathbf{R}\Y_{eq}\Z_g\mathbf{R}^{-1}.
\end{equation}
Now we analyze:
\begin{equation}
\H^t = \I + \L^t = [\mathbf{R}\Y_g + \mathbf{R}\Y_{eq}]\Z_g\mathbf{R}^{-1}.
\end{equation}
Since $\nu_F[\mathbf{R}\Y_{eq},\, \w]+ \nu_F[\mathbf{R}\Y_{g},\, \w]>0$, and $\nu_F[\Z_g\mathbf{R}^{-1},\w]>0$ (which follows from $\nu_F[\mathbf{R}\Y_g,\w]>0$ and the inversion property above, because $\Z_g\mathbf{R}^{-1} = (\mathbf{R}\Y_g)^{-1}$), the same forbidden region applies as for net damping criterion.

Since all specification types share a common forbidden region, the part of the negative real axis to the left of $-1$, $(-\infty, -1]$ (Fig.~\ref{fig:combined_forbidden_regions}), if \eqref{eq:combination} holds the eigenloci of $\L(\jw)$ do not enter this region at any frequency. The eigenloci form closed curves, and the region $(-\infty, -1]$ connects the point $(-1,0)$ to infinity without crossing them; therefore, the eigenloci can't encircle $(-1,0)$. Since $\L(s)$ has no RHP poles, the system is stable according to the generalized Nyquist criterion.
\end{IEEEproof}

It is important to note that stability is \emph{not} guaranteed if different
specification types are assigned to different converters at the same frequency
(e.g., converter~1 satisfies passivity while converter~2 satisfies the
Gershgorin-based specification):
\begin{equation}
\label{eq:combination_wrong}
\forall \w,\ \left[\cnd_1^{dq}(\w) \lor \cnd_1^{\mathrm{P}}(\w)\right] \land \cdots
\land \left[\cnd_n^{dq}(\w) \lor \cnd_n^{\mathrm{P}}(\w)\right] \not\to \text{Stability}.
\end{equation}
A counterexample for this claim is given in the Appendix.
Because of this, the frequency range where a certain type of decentralized specification is satisfied should be provided upfront, as a part of the provided specifications.
For example, combined decentralized specifications can be given as: all converters should satisfy Middlebrook specifications \eqref{eq4:dec-spec-middlebrook-matrix} for frequencies lower than $\w_t$, while for frequencies higher than $\w_t$ they should satisfy passivity-based specifications \eqref{eq4:dec-spec-passivity-matrix}. Other combinations will be presented in subsequent sections.

\subsection{Characterizing the conservativeness of different methods for decentralized specifications}
So far, several methods for imposing decentralized specifications on converters connected to the ac bus have been presented.
All of these methods are frequency-domain conditions.
To compare their conservativeness, for each method and converter $k$ a parameter $D_k(\omega)$ is defined, with $D_k(\omega)>0$ indicating that the specifications are satisfied and $D_k(\omega)<0$ that they are violated.
Since the specifications have different mathematical forms, only the sign of $D_k(\omega)$ is meaningful, and its magnitude cannot be compared across methods; the conservativeness is therefore assessed by evaluating whether the specifications are satisfied over a range of converter parameters, as done in Section~\ref{sec:caseStudies}.
For each type of specification, the parameter $D_k(\w)$ is defined as follows.

Middlebrook (small-gain) based specifications \eqref{eq4:dec-spec-middlebrook-matrix}:
\begin{equation}
D_k^M(\w) = \frac{M_k}{\us\left[\Z_g(\jw)\right]}-\us\left[\Y_k(\jw)\right] .
\label{eq4:DM}
\end{equation}

Passivity based specifications \eqref{eq4:dec-spec-passivity-matrix}:
\begin{equation}
D_k^P(\w) = \nu_F\left[\Y_k,\w\right] + M_k \nu_F\left[\Y_g,\w\right].
\label{eq4:DP}
\end{equation}
Rotated passivity based specifications \eqref{eq4:dec-spec-rotated-passivity-matrix}:
\begin{equation}
D_k^{RP}(\w) = \nu_F\left[\mathbf{R}\Y_k,\w\right] + M_k \nu_F\left[\mathbf{R}\Y_g,\w\right].
\label{eq4:DRP}
\end{equation}

Gershgorin based specifications in $dq$ \eqref{eq:Dec-spec}:
\begin{equation}
\begin{split}
&D_k^{d}(\w) = \Re \big\{L^{dd}_k(\jw) \big\} - \big|L^{dq}_k(\jw)\big| + M_k^d \\
&D_k^{q}(\w) = \Re \big\{L^{qq}_k(\jw) \big\} - \big|L^{qd}_k(\jw)\big| + M_k^q\\
&D_k^{dq}(\w) = \min\left[D_k^{d}(\w),\ D_k^{q}(\w)\right].
\end{split}
\label{eq4:Ddq}
\end{equation}

Gershgorin based specifications in modified sequence domain $pn$ \eqref{eq:Dec-spec_pn}:
\begin{equation}
\begin{split}
&D_k^{p}(\w) = \Re \big\{L^{pp}_k(\jw) \big\} - \big|L^{pn}_k(\jw)\big| + M_k^p \\
&D_k^{n}(\w) = \Re \big\{L^{nn}_k(\jw) \big\} - \big|L^{np}_k(\jw)\big| + M_k^n\\
&D_k^{pn}(\w) = \min\left[D_k^{p}(\w),\ D_k^{n}(\w)\right].
\end{split}
\label{eq4:Dpn}
\end{equation}

%% file: sections/5-Examples.tex
\section{Examples}
\label{sec:caseStudies}

\subsection{Rectifiers connected to the inductive grid}
\label{subsec:exampleA}
\begin{figure}
    \begin{minipage}[b]{0.48\linewidth}
        \centering
        \includegraphics[width = \linewidth]{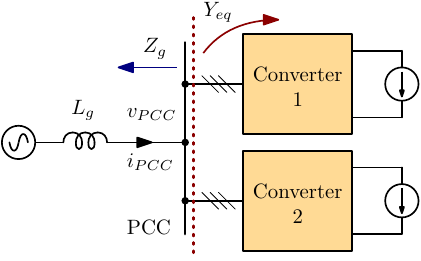}
        \caption{Schematic of the application scenario that shows the two converters planned for connection to the inductive grid (example A).}
        \label{fig:App scenario}
    \end{minipage}
    \hfill
    \begin{minipage}[b]{0.5\linewidth}
        \centering
        \includegraphics[width = \linewidth, page = 2]{figures/simSchm2.pdf}
        \caption{Schematic of the application scenario that shows the two converters planned for connection to the grid with resonance (example B).}
        \label{fig:App scenario2}
    \end{minipage}
\end{figure}

To illustrate the proposed decentralized specifications for ac systems, we consider the symmetric ac grid in Fig.~\ref{fig:App scenario}, with per-phase impedance
$L_g = \qty{0.24}{\pu}$ and $R_{L_g} = \qty{0.06}{\pu}$.
Two two-level, three-phase voltage-source converters are planned for connection to the grid at the PCC.

\begin{figure}[!tb]
    \centering
    \includegraphics[width = 0.6\linewidth]{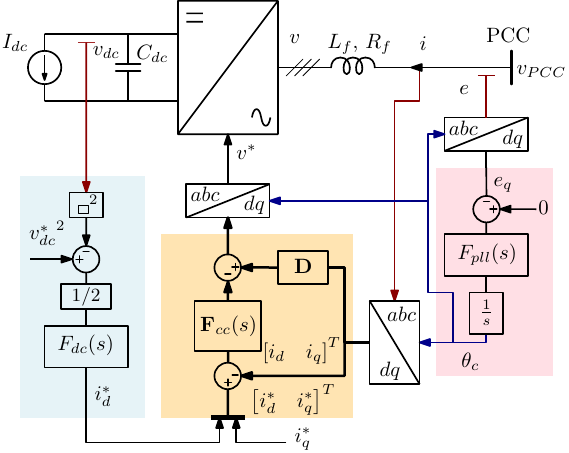}
    \caption{Control structure of the grid-following converters presented in Fig. \ref{fig:App scenario} and \ref{fig:App scenario2}.
    dc-link voltage loop (light blue), current loop (tan), PLL (pink).}
    \label{fig:ConvControl}
\end{figure}

Both converters operate as rectifiers with synchronous-reference-frame control, as shown in Fig.~\ref{fig:ConvControl}.
The parameters for case~I and case~II are listed in Table~\ref{tab:cnv-par}, and the analytical admittance model is taken from~\cite{wenImpedanceBasedAnalysisGridSynchronization2016}.
The power-based scaling factors are chosen as
$M_1 = M_1^d = M_1^q = M_1^p = M_1^n = 2/3$ and
$M_2 = M_2^d = M_2^q = M_2^p = M_2^n = 1/3$.

\begin{table}[!t]
  \renewcommand{\arraystretch}{1.3}
  \caption{Parameters of Example A,  for the \{Case I; Case II\}}
  \label{tab:cnv-par}
  \centering
  \begin{tabular}{l r r}
    \toprule[1 pt]
    \textbf{Parameter} & \textbf{Converter 1} & \textbf{Converter 2} \\
    \midrule
    Steady-state active power $P_0$
                                          & \SInp{2/3}{\pu}      & \SInp{1/3}{\pu} \\
    Steady-state reactive power $Q_0$     & \qty{0}{\pu}      & \qty{0}{\pu} \\
    Steady-state RMS phase current        & \SInp{0.38}{\pu}      & \SInp{0.19}{\pu} \\
    Steady-state $d$ voltage $E_d$
                                          & \SInp{1}{\pu}     & \SInp{1}{\pu} \\
    Steady-state $q$ voltage $E_q$        & \qty{0}{\pu}        & \qty{0}{\pu}     \\
    Dc link capacitance $C_{dc}$
                                            & \qty{1.6}{\pu}       & \qty{1.6}{\pu}  \\
    Filter inductance $L_f$                   & \qty{0.09}{\pu}     & \qty{0.09}{\pu}  \\
   Filter resistance $R_f$                   & \qty{0.036}{\pu}    & \qty{0.036}{\pu}  \\
    Current loop bandwidth                  & \qty{450}{\Hz}         & \qty{700}{\Hz}      \\
    Zero position in current PI  & \SInp{5}{\Hz}  & \SInp{5}{\Hz}      \\
    DVC loop bandwidth                      & \SInp{\{5; 30\}}{\Hz}         & \SInp{\{10; 30\}}{\Hz}      \\
    $K_i/K_p$ of DVC PI controller         &  $\{31; 450\}$   &   $\{63; 450\}$ \\
    PLL loop bandwidth                      & \SInp{\{5; 30\}}{\Hz}    & \SInp{\{10; 30\}}{\Hz}      \\
    $K_i/K_p$ of PLL PI controller            & $\{31; 450\}$   & $\{63; 450\}$ \\
    \bottomrule[1 pt]
\end{tabular}
\end{table}

\begin{figure}[!t]
    \centering
    \includegraphics[page=1]{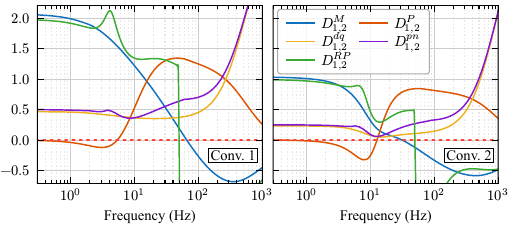}
    \caption{Comparison of different decentralized specifications for converter 1 and 2 from Fig.~\ref{fig:App scenario}, example A, case I.}
    \label{fig5:inductiveGridCnv1}
\end{figure}

In Fig.~\ref{fig5:inductiveGridCnv1} the decentralized specifications for converters~1 and~2 are compared for case~I.
The Gershgorin-based criteria in the $dq$ and modified sequence domains ($D_k^{dq}$, $D_k^{pn}$) are satisfied over the entire frequency range, guaranteeing stability, whereas the small-gain, passivity, and rotated-passivity specifications ($D_k^M$, $D_k^P$, $D_k^{RP}$) are violated in some frequency ranges and are therefore more conservative in this scenario.
Passivity-based specifications $D_k^P$ are violated at low frequencies, while small-gain and rotated-passivity specifications $D_k^M$, $D_k^{RP}$ are violated at higher frequencies.
Rotated passivity $D_k^{RP}$ is satisfied below \qty{50}{\Hz} but exhibits a sharp drop to negative values around \qty{50}{\Hz}, a characteristic behavior also reported in~\cite{chenExtendedFrequencyDomainPassivity2024}.
This complementary behavior motivates combining different specifications across frequency ranges, to obtain specifications that can be satisfied over the entire frequency range.
For the passivity--small-gain combination we define
\begin{equation}
  \label{eq5:combined}
  D^{MP}_k(\omega) =
  \begin{cases}
    D^{P}_k(\omega), & \omega \geq \omega_l, \\
    D^{M}_k(\omega), & \omega < \omega_l ,
  \end{cases}
\end{equation}
and analogously for the passivity--rotated-passivity combination, $D^{RPP}_k(\omega)$,
where $D^{M}_k(\omega)$ is replaced by $D^{RP}_k(\omega)$.

The choice of $\omega_l$ differs between the two combinations.
For $D^{RPP}_k(\omega)$, the sharp drop of $D^{RP}_k(\omega)$ to negative values around \qty{50}{\Hz} provides a natural upper bound, and we select $\omega_l = \qty{47}{\Hz}$ to avoid numerical issues near \qty{50}{\Hz}.
This also exposes a fundamental limitation of the rotated passivity approach: it cannot be used for converters whose bandwidth renders them non-passive near \qty{50}{\Hz}, where $D^{RP}_k(\omega)$ is no longer satisfied.
For $D^{MP}_k(\omega)$, the choice is less constrained, since the passivity and small-gain specifications do not exhibit a distinctive feature around \qty{50}{\Hz}. The value of $\omega_l$ significantly affects conservativeness: for converters with slow dynamics, a lower $\omega_l$ is appropriate, since passivity is then typically satisfied; for faster converters, $\omega_l$ must be higher to avoid relying on passivity at frequencies where it is violated.
However, increasing $\omega_l$ becomes problematic in weak grids, where the small-gain condition is harder to satisfy.
To illustrate this sensitivity, we compare two choices:
$\omega_l = \qty{50}{\Hz}$ (denoted $D^{MP50}$) and $\omega_l = \qty{60}{\Hz}$
(denoted $D^{MP60}$), with $\omega_l = \qty{47}{\Hz}$ also used for $D^{RPP}$.
\begin{figure}[!t]
    \centering
    \includegraphics[page=1]{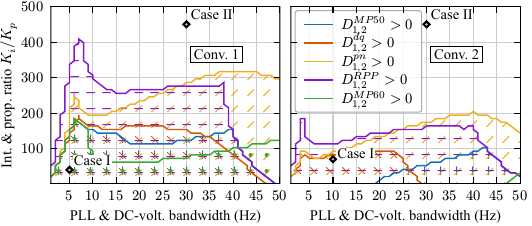}
    \caption{Comparison of satisfaction of decentralized specifications for converters 1 and 2, from Fig.~\ref{fig:App scenario} example A, for different methods.}
    \label{fig5:conservativ}
\end{figure}

To compare the conservativeness of the different specifications, their satisfaction is
evaluated over a parametric space in which the DVC and PLL bandwidths and the corresponding $K_i/K_p$ ratios are varied, as shown in Fig.~\ref{fig5:conservativ}.
The compared specifications are the Gershgorin-based criteria in the $dq$ and modified sequence domains ($D^{dq}$, $D^{pn}$), the passivity--small-gain combinations with $\omega_l = \qty{50}{\Hz}$ and $\omega_l = \qty{60}{\Hz}$ ($D^{MP50}$, $D^{MP60}$),
and the passivity--rotated-passivity combination ($D^{RPP}$).
For both converters, the modified-sequence Gershgorin specification and $D^{RPP}$ are the least conservative.
The $D^{RPP}$ combination is less conservative when the dominant interaction occurs at low frequencies, well below \qty{50}{\Hz}, whereas $D^{pn}$ is less conservative when the interaction shifts toward \qty{50}{\Hz}.
For the passivity--small-gain combinations, increasing $\omega_l$ from \qty{50}{\Hz} to \qty{60}{\Hz} extends the feasible region toward faster dynamics in converter~1, but drastically reduces it to only small areas of the parametric space in converter~2, illustrating that an ill-chosen $\omega_l$ can severely limit the admissible controller parameter set.

\begin{figure}
\begin{minipage}[b]{0.46\linewidth}
    \centering
    \includegraphics[page=1]{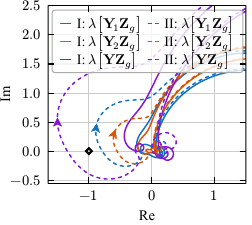}
    \caption{Eigenloci of $\L(\jw)$ ($\w>0$), case~I and case~II, example A}
    \label{fig5:nyquist1}
\end{minipage}\hfill
\begin{minipage}[b]{0.5\linewidth}
    \centering
    \includegraphics[page=1]{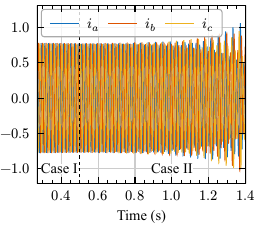}
    \caption{Phase current going to grid $i_{PCC}$ (p.u.), simulation, example A.}
    \label{fig5:inductiveGridTimeDomain}
\end{minipage}
\end{figure}

Finally, stability is verified using the generalized Nyquist criterion applied to three minor-loop gains: $\L_1 = \Y_1\Z_g$ (converter~1 only), $\L_2 = \Y_2\Z_g$ (converter~2 only), and $\L = (\Y_1+\Y_2)\Z_g$ (both converters).
Fig.~\ref{fig5:nyquist1} shows the results for case~I (parameters from Table~\ref{tab:cnv-par}) and case~II (increased DVC and PLL bandwidths and modified $K_i/K_p$ ratios): the system is stable in case~I and unstable in case~II when both converters are connected.

\begin{figure}
    \centering
    \includegraphics[page=1]{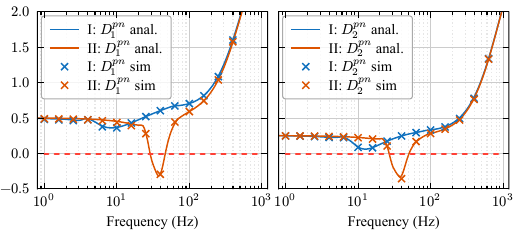}
    \caption{Decentralized specifications for converter 1 and 2, example A, based on Gershgorin theorem in modified sequence domain, for case I and II.}
    \label{fig5:inductiveGridBlackBox}
\end{figure}
To demonstrate that the specifications can be applied directly to black-box measurements, a \emph{Simulink} model is built in which the converter admittances are identified by injecting small-signal perturbations at the output voltage and measuring the resulting input currents, with the converters connected to an ideal grid.
The measured frequency responses are then substituted into~\eqref{eq4:Dpn} to evaluate the decentralized specifications.
Fig.~\ref{fig5:inductiveGridBlackBox} shows the modified-sequence Gershgorin specifications for cases~I and~II, which are satisfied over the entire frequency range in case~I, and violated around \qty{35}{\Hz} in case~II. This is consistent with the analytical results.

The time-domain simulation results are presented in Fig.~\ref{fig5:inductiveGridTimeDomain}, where the controller parameters are switched from case~I to case~II at \qty{0.5}{\second}.
The system becomes unstable after the switch, with oscillations at approximately \qty{35}{\Hz}, consistent with the frequency range of specification violation in Fig.~\ref{fig5:inductiveGridBlackBox}.

\subsection{Grid-feeding inverters connected to the grid with resonance}

In this example, two grid-feeding inverters are connected to a grid with resonance, as shown in Fig.~\ref{fig:App scenario2}, with grid inductance $L_g = \qty{0.15}{\pu}$, capacitance $C_g = \qty{0.05}{\pu}$, and capacitor series resistance $R_{C_g} = \qty{0.1}{\pu}$.
The converter parameters are given in Table~\ref{tab:cnv-par2}, and the scaling factors $M_k$ are power-based, identical to those in example~A.

\begin{table}[!t]
    \renewcommand{\arraystretch}{1.3}
    \caption{Parameters of Example B, for the \{Case I; Case II\}}
    \label{tab:cnv-par2}
    \centering
    \begin{minipage}{\linewidth}
    \begin{tabular}{l r r}
      \toprule[1 pt]
      \textbf{Parameter} & \textbf{Converter 1} & \textbf{Converter 2} \\
      \midrule
      Steady-state active phase power $P_0$
                                            & \SInp{-2/3}{\pu}      & \SInp{-1/3}{\pu} \\
      Steady-state reactive power $Q_0$     & \qty{0}{\pu}      & \qty{0}{\pu} \\
      Steady-state RMS phase current        & \SInp{-0.38}{\pu}      & \SInp{-0.19}{\pu} \\
      Steady-state $d$ voltage $E_d$
                                            & \SInp{1}{\pu}     & \SInp{1}{\pu} \\
      Steady-state $q$ voltage $E_q$        & \qty{0}{\pu}        & \qty{0}{\pu}     \\
      Dc link capacitance $C_{dc}$
                                              & \qty{1.6}{\pu}       & \qty{1.6}{\pu}  \\
     Dc link voltage $E_{dc}$                & \qty{2}{\pu}        & \qty{2}{\pu}  \\
      Filter inductance $L_f$                   & \qty{0.03}{\pu}     & \qty{0.06}{\pu}  \\
     Filter resistance $R_f$                   & \qty{0.005}{\pu}    & \qty{0.005}{\pu}  \\
      Current loop bandwidth                  & \qty{1}{\kHz}         & \qty{1}{\kHz}      \\
      Zero position in current PI             & \SInp{20}{\Hz}  & \SInp{20}{\Hz}      \\
      Delay present in the control loop       & \SInp{\{100; 200\}}{\mu s} & \SInp{\{100; 200\}}{\mu s} \\
      DVC loop bandwidth                      & \qty{20}{\Hz}         & \qty{20}{\Hz}      \\
      Ratio of DVC PI controller $K_i/K_p$                       & \qty{125}{}   & \qty{125}{} \\
      PLL loop bandwidth                      & \qty{20}{\Hz}         & \qty{20}{\Hz}      \\
      Ratio of PLL PI controller $K_i/K_p$                     & \qty{125}{}   & \qty{125}{} \\
      \bottomrule[1 pt]
  \end{tabular}
  \end{minipage}
  \end{table}

\begin{figure}[!t]
    \centering
        \includegraphics[page=2]{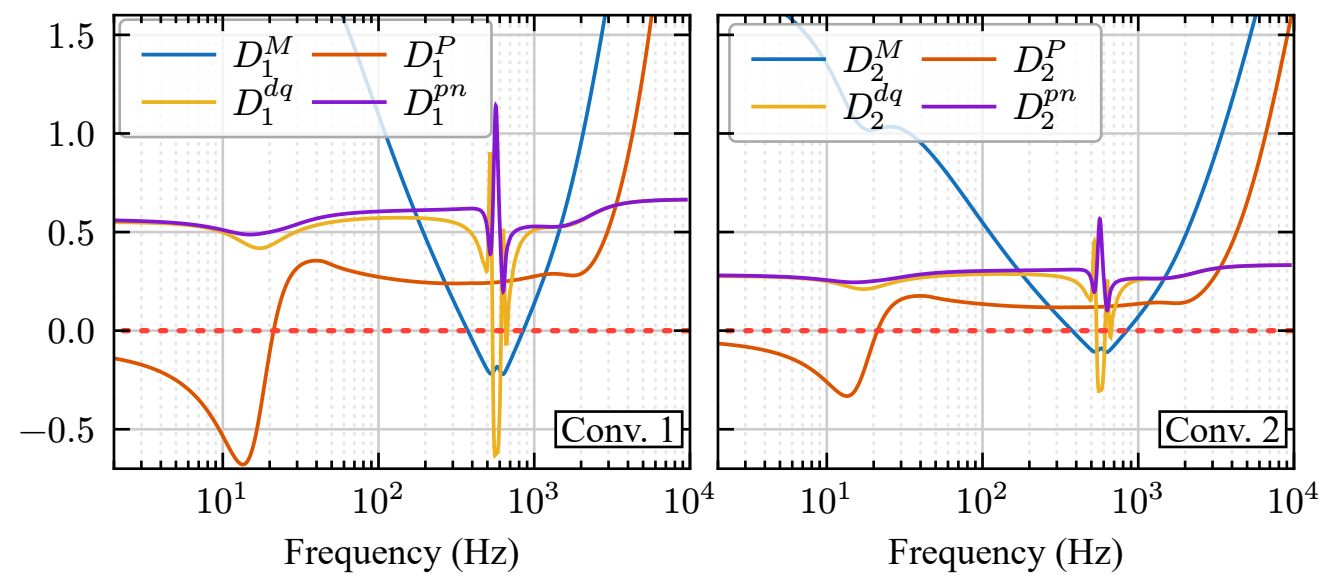}
        \caption{Comparison of different decentralized specifications for converter 1 and 2 from Fig.~\ref{fig:App scenario2}, example B, case I.}
        \label{fig5:resGridCnv1}
\end{figure}

Fig.~\ref{fig5:resGridCnv1} shows that only the modified-sequence Gershgorin specification is satisfied over the entire frequency range, while all other specifications are violated in some ranges.
Notably, passivity is satisfied precisely where the small-gain and $dq$ Gershgorin specifications fail, motivating a combination of the two to achieve full-range coverage.
Rotated passivity is not considered here, as the interactions of interest occur above \qty{50}{\Hz}, where $D_k^{RP}$ is not satisfied.

\begin{figure}[!t]
    \centering
    \includegraphics[page=2]{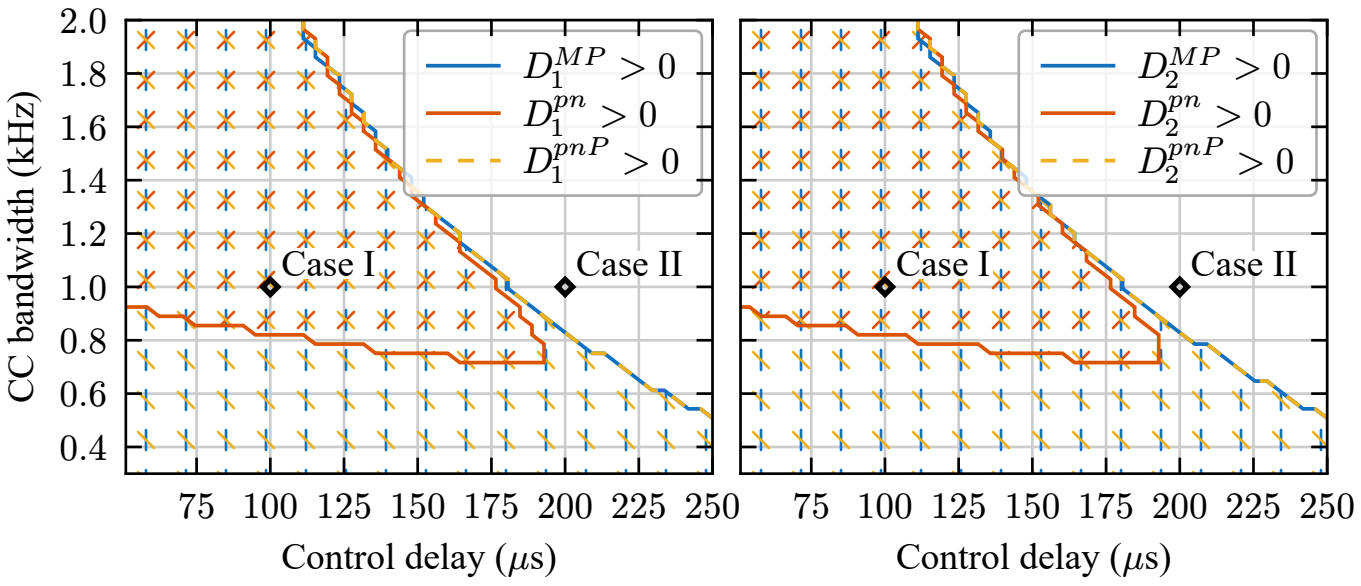}
    \caption{Comparison of satisfaction of decentralized specifications for converter 1 and 2, from Fig.~\ref{fig:App scenario2} example B, for different methods.}
    \label{fig5:resConservativ}
\end{figure}

To compare the conservativeness of the different methods, specification satisfaction is evaluated in a parametric space of current-loop bandwidth and control delay, as shown in Fig.~\ref{fig5:resConservativ}.
The compared specifications are the modified-sequence Gershgorin criterion $D^{pn}$, the passivity--small-gain combination $D^{MP}$, and the passivity--Gershgorin combination $D^{pnP}$, defined as:
\begin{equation}
  \label{eq5:combined2}
  D^{pnP}_k(\omega) =
  \begin{cases}
    D^{P}_k(\omega),  & \omega \in [\qty{50}{\Hz},\, \qty{2}{\kHz}], \\
    D^{pn}_k(\omega), & \omega \in (0, \qty{50}{\Hz}) \cup (\qty{2}{\kHz}, \infty),
  \end{cases}
\end{equation}
where passivity governs the mid-frequency range where the small-gain and Gershgorin $dq$ specifications are violated.

From Fig.~\ref{fig5:resConservativ}, the standalone modified-sequence Gershgorin specification $D^{pn}$ is satisfied only for a limited parameter set.
When combined with passivity, both $D^{MP}$ and $D^{pnP}$ yield comparable conservativeness, indicating that the small-gain and modified-sequence Gershgorin bases perform similarly in this scenario.
Analogous results hold for the $dq$ Gershgorin criterion but are omitted from the figure for readability.

\begin{figure}[!t]
    \begin{minipage}[b]{0.47\linewidth}
        \centering
        \includegraphics[page=2]{nyquist.pdf}
        \caption{Eigenloci of $\L(\jw)$ for case~I and case~II, example B.}
        \label{fig5:nyquist2}
    \end{minipage}\hfill
    \begin{minipage}[b]{0.5\linewidth}
        \centering
        \includegraphics[page=2]{simTime.pdf}
        \caption{Phase current going to the grid $i_{PCC}$ (p.u.), simulation, example~B.}
        \label{fig5:resGridTimeDomain}
    \end{minipage}
\end{figure}

Applying the generalized Nyquist criterion to the system in Fig.~\ref{fig:App scenario2} with case~I parameters confirms stability, as there are no encirclements of the $(-1,0)$ point (Fig.~\ref{fig5:nyquist2}), consistent with the satisfied specifications in Fig.~\ref{fig5:resConservativ}.
Increasing the control delay to \qty{200}{\mu s} (case~II) renders the system unstable when both converters are connected, as confirmed by the Nyquist plot.
Since instability occurs almost immediately upon leaving the specification-satisfied region, the combined Gershgorin--passivity specifications introduce negligible conservativeness in this example.

\begin{figure}
    \centering
    \includegraphics[page=3]{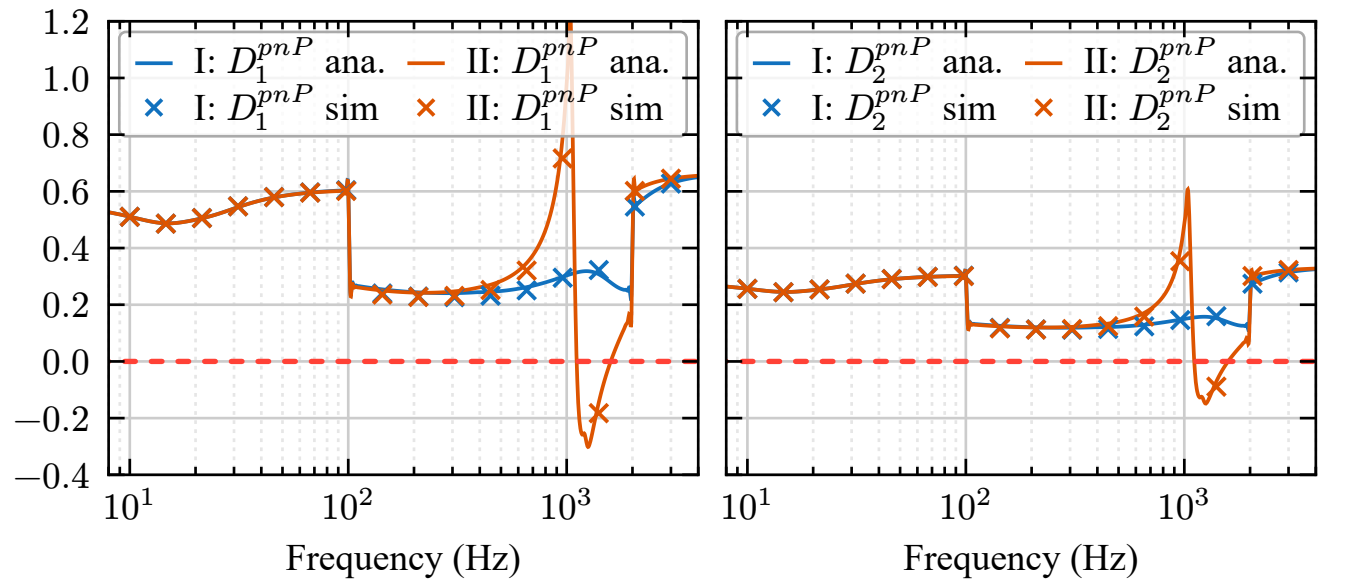}
    \caption{Decentralized specifications for converter 1 and 2, example B, based on combination of Gershgorin theorem in modified sequence domain and passivity, $D^{pnP}$, for case I and II.}
    \label{fig5:resGridBlackBox}
\end{figure}

As in example~A, a \emph{Simulink} model is used for small-signal measurements of converter admittance and grid impedance.
The combined Gershgorin--passivity specification~\eqref{eq5:combined2} is then evaluated from the measurements and shown for converter~1 and 2 in Fig.~\ref{fig5:resGridBlackBox}: satisfied over the entire frequency range in case~I, and violated around \qty{1.5}{\kHz}
in case~II.

The time-domain simulation results are shown in Fig.~\ref{fig5:resGridTimeDomain}, where the controller parameters are switched from case~I to case~II at \qty{0.5}{\second}.
The system becomes unstable after the switch, with oscillations at approximately \qty{1.5}{\kHz}, consistent with the frequency range of specification violation in Fig.~\ref{fig5:resGridBlackBox}.

\subsection{Grid-forming converters connected to the inductive grid}
\label{subsec:exampleC}
\begin{figure}[!tb]
    \centering
    \includegraphics[width = 0.9\linewidth]{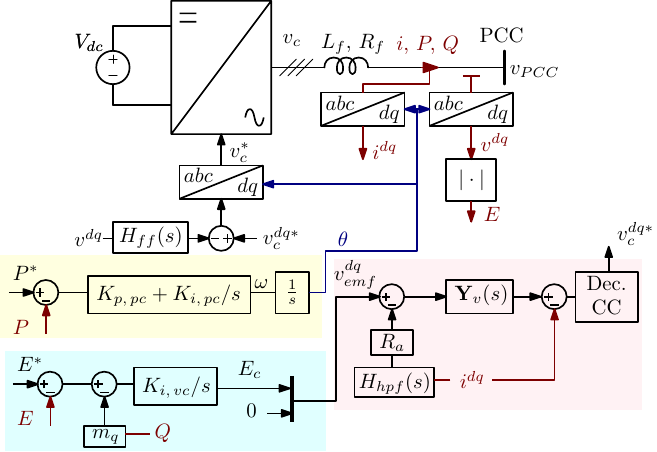}
    \caption{Control structure of the grid-forming converters, example C.}
    \label{fig:GFMControl}
\end{figure}

Up to now, the proposed specifications have been tested on grid-following converters.
In this example, two grid-forming converters with active power synchronization, voltage magnitude control, droop on reactive power for parallel operation, and virtual-admittance control structure shown in Fig.~\ref{fig:GFMControl} are connected to the PCC of an inductive grid with inductance $L_g = \qty{0.1}{\pu}$ and resistance $R_{L_g} = \qty{0.02}{\pu}$, like in Fig.~\ref{fig:App scenario}.
The converter parameters are listed in Table~\ref{tab:cnv-par3}, where all per-unit values refer to the global base, i.e., the total rated power of both converters; case~II triples the voltage-loop and power-loop bandwidths of both converters.
The scaling factors $M_k$ are power-based, identical to those in example~A.

\begin{table}[!t]
  \renewcommand{\arraystretch}{1.3}
  \caption{Parameters of Example C, for the \{Case I; Case II\}}
  \label{tab:cnv-par3}
  \centering
  \begin{minipage}{\linewidth}
  \centering
  \begin{tabular}{l r r}
    \toprule[1 pt]
    \textbf{Parameter}\footnote{All per-unit values are referred to the global base, i.e., the total rated power of both converters. Therefore, some parameters may appear unrealistically large; e.g., since converter~2 has a rated power of \SInp{1/3}{\pu}, its virtual inductance $L_v = \qty{1.41}{\pu}$ corresponds to $1.41/3 = \qty{0.47}{\pu}$ on its own rating, i.e., in the per-unit system where $S_2 = \qty{1}{\pu}$.} & \textbf{Converter 1} & \textbf{Converter 2} \\
    \midrule
    Rated power $S_k$                      & \SInp{2/3}{\pu}      & \SInp{1/3}{\pu} \\
    Steady-state active power $P_0$         & \qty{0.56}{\pu}      & \qty{0.28}{\pu} \\
    Steady-state reactive power $Q_0$       & \qty{0}{\pu}         & \qty{0}{\pu} \\
    Filter inductance $L_f$                 & \qty{0.075}{\pu}     & \qty{0.15}{\pu} \\
    Filter resistance $R_f$                 & \qty{0.0075}{\pu}    & \qty{0.015}{\pu} \\
    Virtual inductance $L_v$                & \qty{0.705}{\pu}     & \qty{1.41}{\pu} \\
    Virtual resistance $R_v$                & \qty{0.465}{\pu}     & \qty{0.93}{\pu} \\
    Current loop bandwidth                  & \qty{500}{\Hz}       & \qty{500}{\Hz} \\
    Voltage loop bandwidth                  & \SInp{\{0.7; 2.1\}}{\Hz} & \SInp{\{1; 3\}}{\Hz} \\
    $Q$--$V$ droop $m_q$                    & \qty{0.18}{\pu}      & \qty{0.3}{\pu} \\
    Transient damping resistance $R_{a,vc}$ & \qty{0.63}{\pu}      & \qty{1.5}{\pu} \\
    Transient damping high-pass filter      & \qty{5.3}{\Hz}       & \qty{4}{\Hz} \\
    Power loop bandwidth                    & \SInp{\{9.5; 28.5\}}{\Hz} & \SInp{\{9.5; 28.5\}}{\Hz} \\
    Active damping ratio $R_a/K_{p,pc}$     & \qty{1.1}{}          & \qty{1.4}{} \\
    \bottomrule[1 pt]
  \end{tabular}
  \end{minipage}
\end{table}

\begin{figure}[!t]
    \centering
    \includegraphics[page=3]{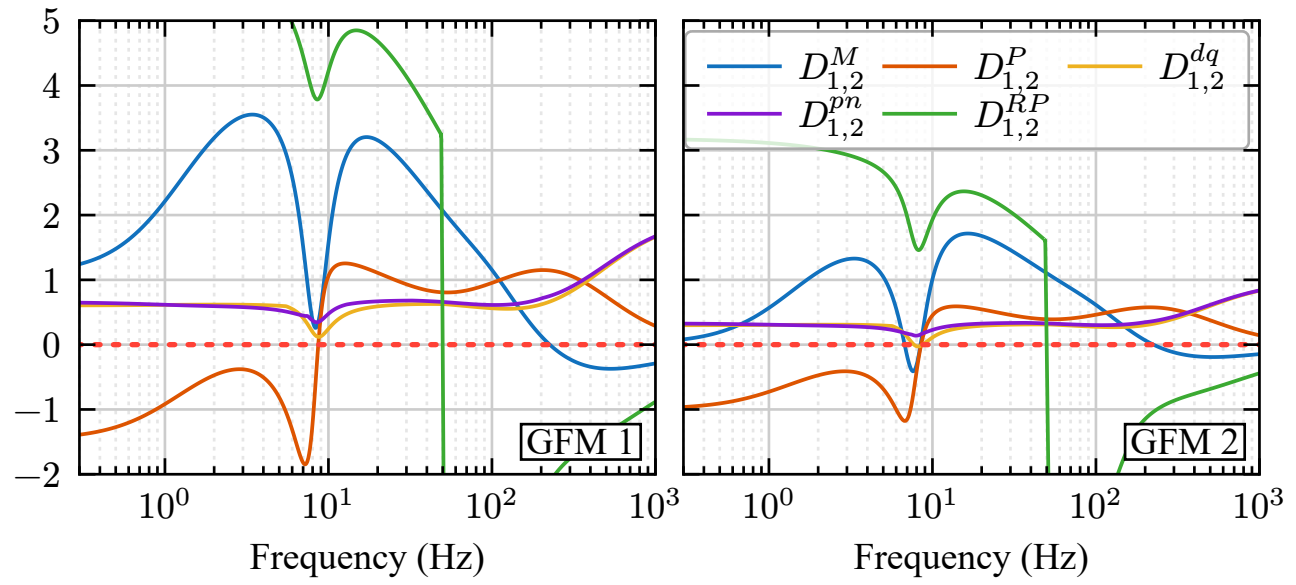}
    \caption{Comparison of different decentralized specifications for grid-forming converters 1 and 2, example C, case I.}
    \label{fig5:gfmSpecs}
\end{figure}

Fig.~\ref{fig5:gfmSpecs} compares the decentralized specifications for case~I.
The modified-sequence Gershgorin specification $D_k^{pn}$ is satisfied over the entire frequency range for both converters, whereas $D_k^{dq}$ is violated for converter~2 around \qty{8}{\Hz}.
In contrast to the grid-following converters, the grid-forming converters are non-passive at low frequencies, below approximately \qty{9}{\Hz}, where the power loop is active, while small gain is violated above approximately \qty{230}{\Hz}, and additionally around \qty{8}{\Hz} for converter~2.
Rotated passivity is satisfied below \qty{50}{\Hz}, which covers the frequency range where passivity fails; hence, the combination $D_k^{RPP}$ is satisfied for both converters, while $D_k^{MP50}$ is violated for converter~2.

\begin{figure}[!t]
    \centering
    \includegraphics[page=3]{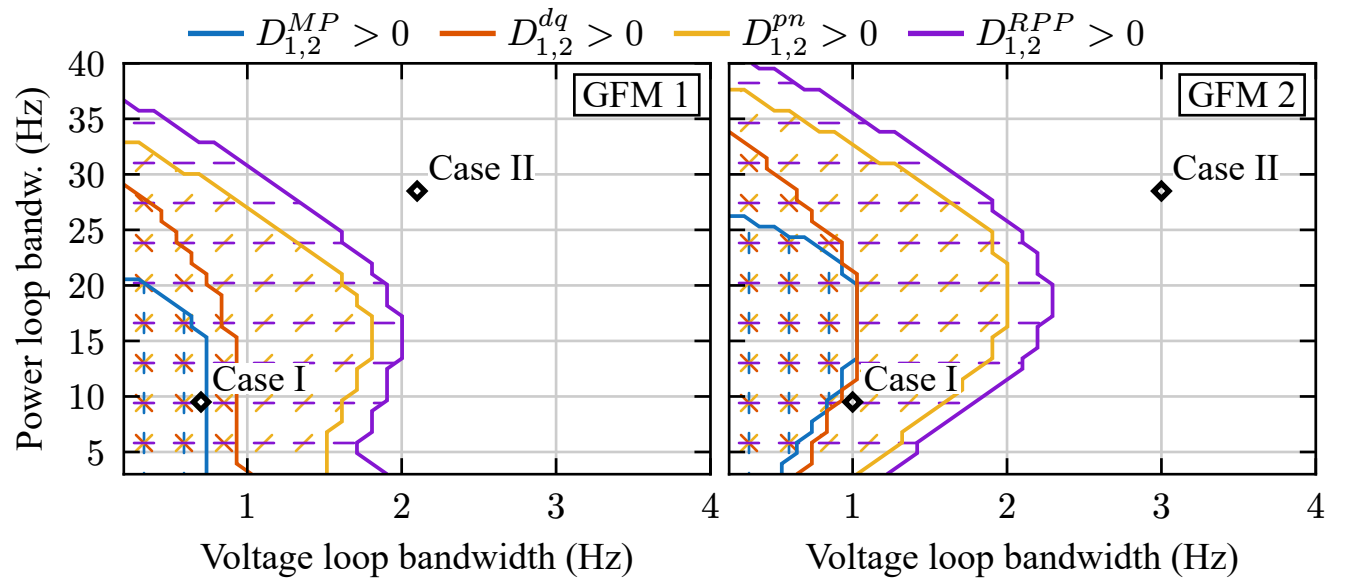}
    \caption{Comparison of satisfaction of decentralized specifications for grid-forming converters 1 and 2, example C, for different methods. For each map, only the loops of the corresponding converter are varied.}
    \label{fig5:gfmConservativ}
\end{figure}

The conservativeness of the specifications is compared in a parametric space of voltage-loop and power-loop bandwidths, shown in Fig.~\ref{fig5:gfmConservativ}.
As in example~A, $D^{pn}$ and $D^{RPP}$ are the least conservative and yield regions of comparable size, followed by $D^{dq}$, while $D^{MP50}$ is the most conservative, since small gain must cover the low-frequency range where the power-loop dynamics make the converters non-passive.
The case~I parameters lie inside the regions where $D^{pn}$ and $D^{RPP}$ are satisfied for both converters, whereas the case~II parameters lie outside all regions.

\begin{figure}[!t]
    \begin{minipage}[b]{0.47\linewidth}
        \centering
        \includegraphics[page=3]{nyquist.pdf}
        \caption{Eigenloci of $\L(\jw)$ for case~I and case~II, example C.}
        \label{fig5:nyquist3}
    \end{minipage}\hfill
    \begin{minipage}[b]{0.5\linewidth}
        \centering
        \includegraphics[page=3]{simTime.pdf}
        \caption{Phase current going to the grid $i_{PCC}$ (p.u.), simulation, example~C.}
        \label{fig5:gfmTimeDomain}
    \end{minipage}
\end{figure}

The generalized Nyquist criterion in Fig.~\ref{fig5:nyquist3} confirms that the system is stable in case~I.
In case~II, each converter is stable when connected alone ($\L_1$ and $\L_2$ do not encircle $(-1,0)$), but the system with both converters is unstable, with eigenloci of $\L$ encircling $(-1,0)$ around \qty{23}{\Hz}.
Thus, the instability results from the interaction of the two converters, and it is captured by the decentralized specifications, even though each converter alone would pass a conventional single-converter stability assessment.

\begin{figure}
    \centering
    \includegraphics[page=4]{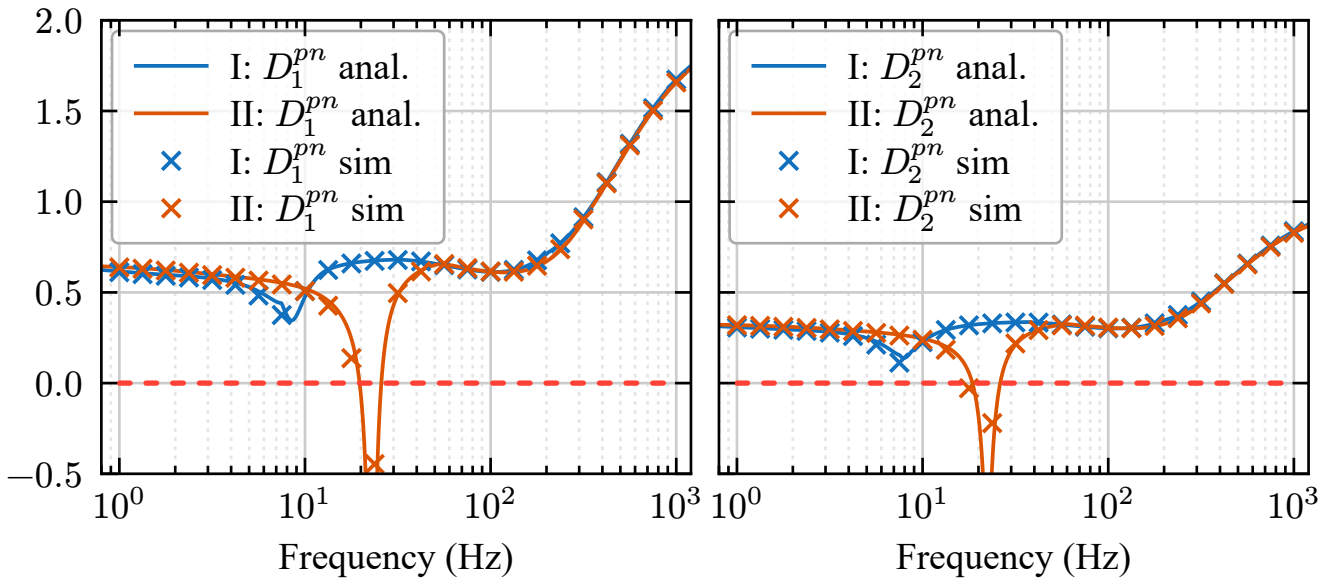}
    \caption{Decentralized specifications for grid-forming converters 1 and 2, example C, based on Gershgorin theorem in modified sequence domain, for case I and II.}
    \label{fig5:gfmBlackBox}
\end{figure}

As in the previous examples, the converter admittances are identified from small-signal measurements in \emph{Simulink}, and the modified-sequence Gershgorin specifications are evaluated from the measurements, as shown in Fig.~\ref{fig5:gfmBlackBox}.
The specifications are satisfied over the entire frequency range in case~I and violated approximately between \qty{20}{\Hz} and \qty{30}{\Hz} for both converters in case~II, in agreement with the analytical model.

The time-domain simulation results are shown in Fig.~\ref{fig5:gfmTimeDomain}, where the controller parameters are switched from case~I to case~II at \qty{1}{\second}.
The system becomes unstable after the switch, with oscillations at approximately \qty{23}{\Hz}, consistent with the frequency range of specification violation in Fig.~\ref{fig5:gfmBlackBox}.

\subsection{Mixed grid-forming and grid-following converters connected to the inductive grid}
\label{sec:mixed}

Since the specifications depend only on the frequency response of each converter, they apply equally to systems combining different control strategies.
To illustrate this, converter~2 in example~C is replaced by the grid-following converter~2 from example~A.
In case~II, only the grid-forming converter~1 is switched to the case~II parameters from Table~\ref{tab:cnv-par3}, while the grid-following converter~2 is left unchanged.

\begin{figure}[!t]
    \centering
    \includegraphics[width=\linewidth]{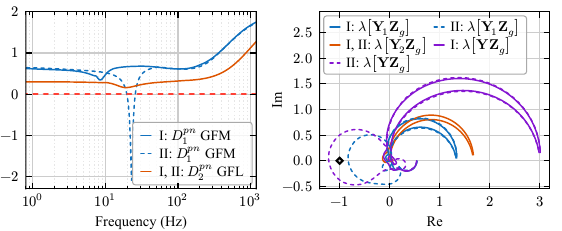}
    \caption{Mixed grid-forming converter~1 and grid-following converter~2, cases~I and~II: decentralized specifications in the modified sequence domain (left) and eigenloci of $\L(\jw)$ (right).}
    \label{fig5:mixed}
\end{figure}

As shown in Fig.~\ref{fig5:mixed}, the specifications are satisfied for both converters in case~I, whereas in case~II they are violated around \qty{23}{\Hz} by converter~1 only.
This single violation is sufficient to destabilize the system, as confirmed by the eigenloci of $\L$ encircling $(-1,0)$, even though converter~1 alone remains stable.

\section{Experimental Results}
\label{sec:experimental}

\begin{figure}
    \centering
    \includegraphics[page=1, width = 0.8\linewidth]{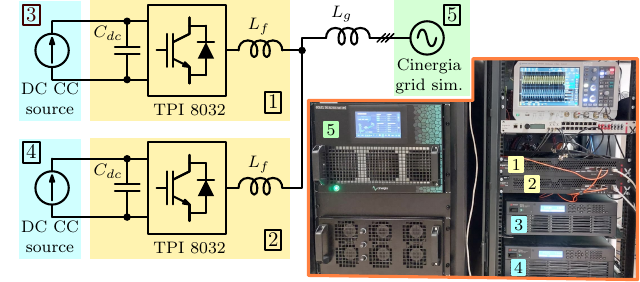}
    \caption{Experimental setup.}
    \label{fig:ExpSetup}
\end{figure}

So far, the specifications have been applied to analytical models and simulation measurements. One of the key benefits of decentralized specifications is that, since they require only the frequency response of each individual converter, they can be evaluated directly from measurements on real devices, without a converter analytical or simulation model. To demonstrate this, example~A is implemented on the experimental setup shown in Fig.~\ref{fig:ExpSetup}, consisting of two three-phase \emph{Imperix} TPI~8032 inverters connected to a \emph{Cinergia} grid emulator through an inductive impedance, with each inverter dc side supplied by a \emph{Keysight} power supply in constant-current mode.
The system parameters match Table~\ref{tab:cnv-par}, with base values of \qty{100}{\V} (voltage) and \qty{3}{kW} (power). The inverters switch at \qty{50}{kHz}, with the control executed once per switching period.

\begin{figure}[!tp]
    \centering
    \includegraphics[page=2, width = \linewidth]{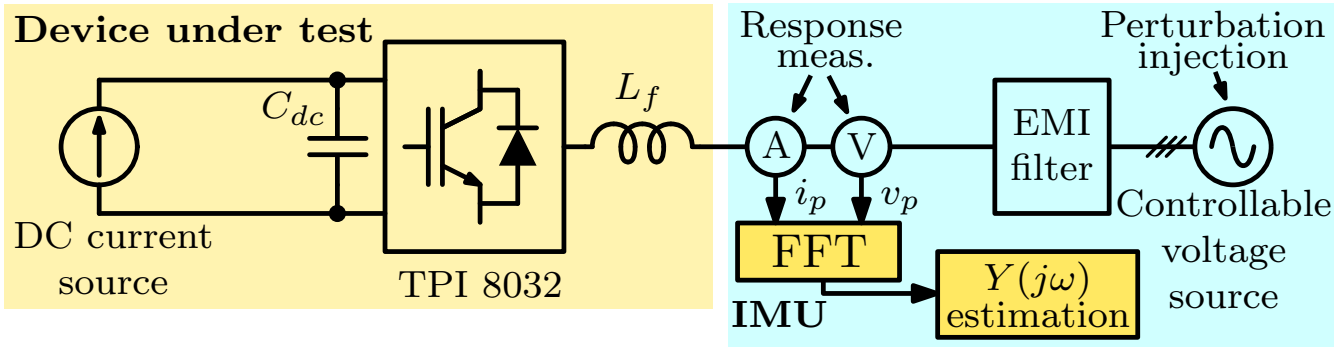}
    \caption{Schematic representing the procedure for measuring the converter admittance.}
    \label{fig:ExpSetupImpMeas}
\end{figure}

The converter admittance is identified using an impedance measurement unit (IMU) implemented on an additional \emph{Imperix} TPI~8032 inverter, whose dc side is connected to a constant-voltage source and whose ac output modulates a small-signal perturbation on top of the steady-state fundamental, as shown in Fig.~\ref{fig:ExpSetupImpMeas}. An EMI filter suppresses switching noise.
With the system in steady state, a perturbation at frequency $\omega_p$ is injected,
the phase voltages $v_p$ and currents $i_p$ are measured, transformed to the $dq$
frame, and the admittance $Y(j\omega_p)$ is obtained via Fourier analysis.
This procedure is repeated for both converters in cases~I and~II.

\begin{figure}[!t]
    \centering
    \includegraphics[page=2]{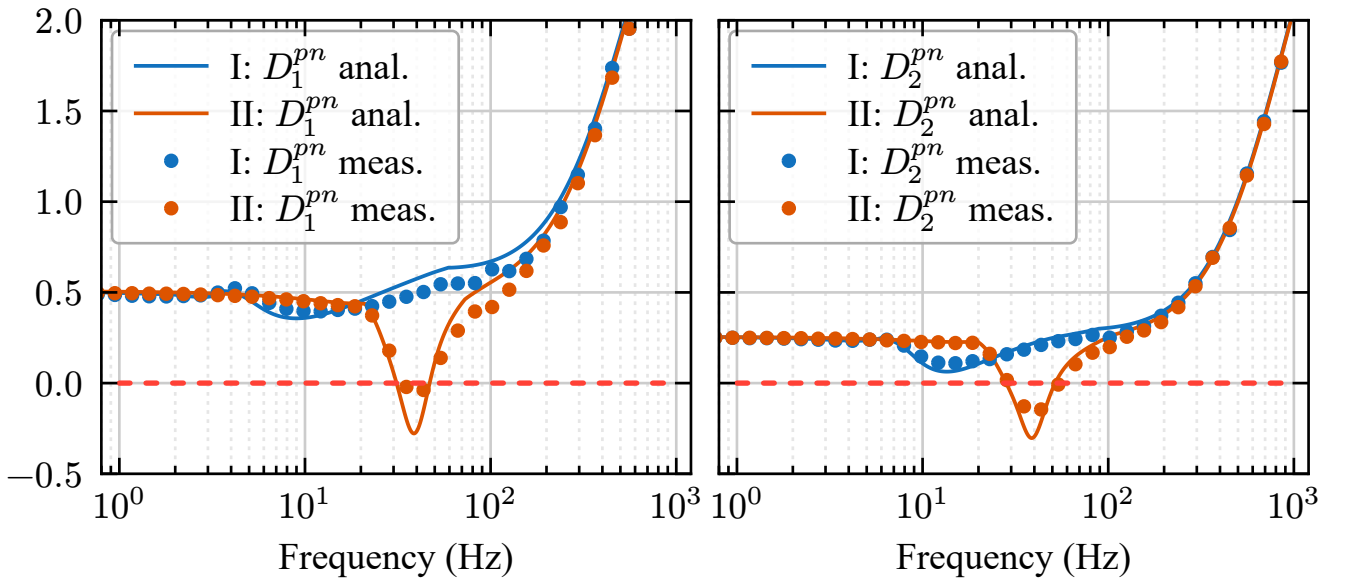}
    \caption{Experimental results of specifications in the modified sequence domain for case I and II, example A.}
    \label{fig:ExpSpecs}
\end{figure}

\begin{figure}[!t]
    \centering
    \includegraphics[]{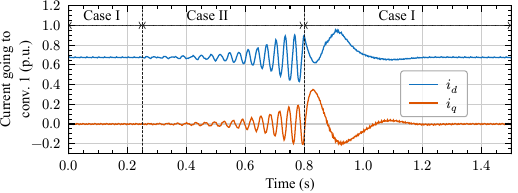}
    \caption{Experimental results, $i_d$ and $i_q$ from converter 1, case I and II, example A.}
    \label{fig:ExpTime}
\end{figure}

\begin{figure}[!t]
    \centering
    \includegraphics[width = 0.9\linewidth]{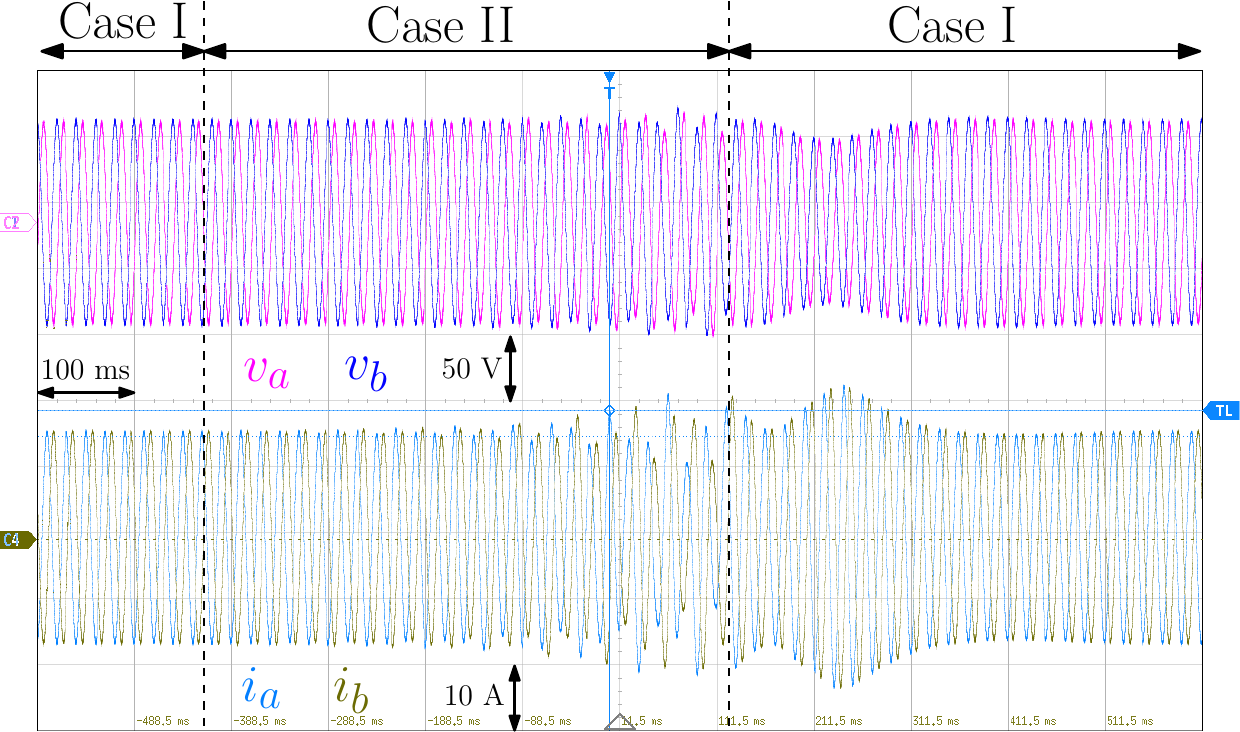}
    \caption{Change of the parameters from case I to case II for phase voltages and currents of converter 1, results from oscilloscope.}
    \label{fig:ExpScope}
\end{figure}

The decentralized specifications are then evaluated from the measured admittances, and the results are shown in Fig.~\ref{fig:ExpSpecs}: the modified-sequence Gershgorin specifications are satisfied over the entire frequency range in case~I and violated around \qty{35}{\Hz} in case~II, in agreement with both the analytical model and simulation.
Time-domain experimental results are presented in Fig.~\ref{fig:ExpTime} (converter~1 internal signals) and Fig.~\ref{fig:ExpScope} (oscilloscope waveforms). Switching the parameters from case~I to case~II causes instability with oscillations at approximately \qty{35}{\Hz}, confirming that specification violation in a given frequency range can lead to system instability.

%% file: sections/6-Conclusions.tex
\begin{table*}
\caption{Summary of the applicability of different methods for different frequency ranges.}
\label{tab:summary}
\centering
\begin{tabular}{l c c c}
    \toprule
    \textbf{Method} & \textbf{Low-frequency subsynchronous range} & \textbf{Synchronous range} & \textbf{Harmonic range} \\
    \midrule
    Small-gain theorem                  & Usable        & Usable, but conservative & Not applicable \\
    Gershgorin theorem in $dq$ domain    & Good          & Good                    & Applicable only without resonances \\
    Gersh. in modified sequence domain  & Very good      & Very good                      & Applicable if resonances are damped \\
    Passivity-based                     & Not applicable & Very good - in most cases                  & Excellent \\
    Rotated passivity-based            & Excellent & Not applicable           & Not applicable \\
    \bottomrule
\end{tabular}
\end{table*}

\section{Discussion}
Across all examples, covering GFL converters, GFM converters, and their combination, as well as sub-synchronous and harmonic-range interactions, two observations hold.
First, the specifications do not lead to an overly conservative design; some conservativeness is inherent, since each converter's specification is independent of the others, but this is the price of decentralized design.
Second, when the specifications are violated by changing the converter parameters, the system can become unstable. In the mixed example of Section~\ref{sec:mixed}, a violation by a single converter, which is itself stable when connected alone, can also lead to system instability.

The results presented in Sections~\ref{sec:caseStudies} and~\ref{sec:experimental} also show that the most suitable type of decentralized specification depends on the frequency range of interest. The frequency ranges can be broadly classified as follows.
\begin{itemize}
\item \textbf{Low-frequency subsynchronous range} (up to tens of \unit{\Hz}): In this range converters fulfill their primary function of supplying active and reactive power, and are typically not passive; net-damping specifications are therefore almost inapplicable. Small-gain, Gershgorin-based ($dq$ and modified sequence domain), and rotated-passivity specifications are the most suitable. In the tested scenarios, the
least conservative specifications are those based on rotated passivity, followed by the Gershgorin criterion in the modified sequence domain and then in the $dq$ domain. Small-gain-based specifications are also applicable, but tend to be more conservative.

\item \textbf{Synchronous range} (frequencies around the fundamental frequency): The $d$--$q$ coupling is pronounced in this range. For converters with sufficiently slow dynamics, which is typical in power-system applications, passivity-based specifications are applicable; for converters with higher control bandwidths, passivity cannot be used. Rotated passivity is inapplicable in this range because $D_k^{RP}(\omega)$ drops sharply to negative values around the fundamental frequency, as discussed in Section~\ref{subsec:exampleA}. Gershgorin-based specifications are well-suited here, with the modified-sequence-domain variant being less conservative owing to its inherent handling of $d$--$q$ coupling. The small-gain theorem is also applicable but generally more conservative.

\item \textbf{Harmonic range} (from hundreds of \unit{\Hz} to a few \unit{\kHz}): Converters are not expected to exchange active or reactive power in this range and should be passive. Passivity-based specifications are therefore the most suitable and robust choice, as demonstrated in example~B. In scenarios without undamped resonances, Gershgorin-based specifications can also be applied, though passivity-based specifications are preferable in general.

\item \textbf{High-frequency range} (above a few \unit{\kHz}): This range exceeds the control bandwidth of the converter, where modulation, delays, and other non-idealities may cause the converter to be non-passive. When EMI filters render the converter passive at these frequencies, passivity-based specifications apply; otherwise, Gershgorin-based or small-gain specifications may be used. Note that this range is not covered by the examples presented and is included here for completeness; it is generally of greater relevance to single-converter design than to multi-converter grid stability.
\end{itemize}
This classification (except for the high-frequency range, which is not of interest for this paper) is summarized in Table~\ref{tab:summary}.

Specification providers can therefore tailor the specifications to the grid the converters connect to. For mainly inductive grids without many undamped resonances, Gershgorin-based specifications in the modified sequence domain suffice over the whole frequency range. When harmonic-range issues are expected, passivity-based specifications are preferable. Since passivity must be combined with other specifications, the frequency range in which the converter must be passive should be stated. Its lower bound depends on the desired control speed and typically lies in the 10--40~Hz range; below it, rotated-passivity or Gershgorin-based specifications apply. If control delays are expected to cause active behavior at higher frequencies, an upper bound should also be set (as in example~B), with small-gain or Gershgorin-based specifications above it.

Finally, since the specifications are based on small-signal impedances, they hold only at the operating point at which the impedance is obtained. Connecting new converters may shift the operating point of the existing ones (e.g., the PCC voltage or the power flow), so each converter should satisfy its specification over its whole operating range. As this range is known upfront from the converter ratings and grid-code requirements, compliance can be verified analytically or by impedance measurements at a set of representative operating points, such as the corners of the operating range.
\section{Conclusions}

This paper presents a systematic treatment of decentralized impedance specifications for power electronics converters connected to a common ac bus. The following contributions are made.
First, decentralized specifications based on the small-gain theorem, passivity, and the Gershgorin theorem are unified in a single framework, with simple closed-form frequency-domain expressions provided for each method. As a novel extension of the Gershgorin-based approach, decentralized specifications formulated in the modified sequence domain are proposed, which reduce the conservativeness introduced by the
Gershgorin theorem in the $dq$ domain by exploiting the near-diagonal structure of symmetric impedances in the modified sequence domain.
Second, a general framework for combining different types of decentralized specifications across distinct frequency ranges is introduced. Using the forbidden region principle, it is proved that the combined specifications remain a sufficient condition for system stability, provided that, at each frequency, the same type of specification is satisfied by all converters. Assigning different specification types to different converters at the same frequency does not guarantee stability, as shown by the counterexample in the Appendix.
Third, the conservativeness and applicability of the different methods are systematically compared on three examples: sub-synchronous interactions of GFL rectifiers connected to an inductive grid (example~A), harmonic-range interactions of GFL inverters connected to a resonant grid (example~B), and sub-synchronous interactions of GFM converters connected to an inductive grid (example~C); their applicability to systems combining different control strategies is further demonstrated on a mixed GFM--GFL system.
Based on these results, practical guidelines are provided for selecting the most appropriate specification type depending on the frequency range and scenario:
rotated-passivity and Gershgorin-based methods are preferred at sub-synchronous frequencies,
Gershgorin-based methods are most suitable around the fundamental frequency,
and passivity-based specifications are the natural choice in the harmonic range.
Finally, the specifications are applied directly to black-box frequency-response
measurements obtained from both simulation and a laboratory experimental setup,
without relying on analytical converter models. It is consistently shown that the proposed specifications are a valuable tool for preventing unwanted interactions and ensuring stability in multi-converter ac systems.

%% file: sections/7-Appendix.tex
\section*{Appendix}

This appendix provides a counterexample showing that, at a single frequency
$\omega_c$, combining different types of decentralized specifications across
converters does not necessarily guarantee stability, as discussed in
\eqref{eq:combination_wrong}. Since we focus on a single frequency, only the
impedance and admittance values at $\omega_c$ are considered.

Consider two converters of equal nominal power, so that the power-based scaling factors are $M_1 = M_2 = 0.5$ and $\sum_k M_k = 1$. The same values are used for all specification types, i.e., $M_k^d = M_k^q = M_k$.
Let the admittances of the converters and the grid impedance at $\omega_c$ be given by:
\begin{equation}
\begin{split}
    \Y_1 = \begin{bmatrix}
        -2j & 0 \\
        0 & 1
    \end{bmatrix},&\quad
    \Y_2 = \begin{bmatrix}
        -1+j & 0 \\
        0 & 1
    \end{bmatrix} \\
    \Z_g &= \begin{bmatrix}
        1-j & 0 \\
        0 & 1
    \end{bmatrix}
\end{split}
\end{equation}

The grid is passive, since $\Y_g = \Z_g^{-1} = \mathrm{diag}(0.5+0.5j,\ 1)$ gives $\nu_F[\Y_g,\w_c] = 0.5 > 0$. The minor loop gains of the converters are:
\begin{equation}
    \L_1 = \Y_1\Z_g = \begin{bmatrix}
        -2-2j & 0 \\
        0 & 1
    \end{bmatrix},\quad
    \L_2 = \Y_2\Z_g = \begin{bmatrix}
        2j & 0 \\
        0 & 1
    \end{bmatrix}.
\end{equation}
Now we can calculate the specifications based on passivity and Gershgorin theorem in $dq$ for converters 1 and 2, according to \eqref{eq4:DP} and \eqref{eq4:Ddq}:
\begin{equation}
    \begin{split}
    &D_1^{P}(\w_c) = 0.25, \ \ \ \ \ D_1^{dq}(\w_c) = -1.5, \\
    &D_2^{P}(\w_c) = -0.75, \ \ \ D_2^{dq}(\w_c) = 0.5.
    \end{split}
\end{equation}
From these values, we can see that converter 1 satisfies the passivity-based specification, while converter 2 satisfies the Gershgorin-based specification in $dq$ domain. However, the loop gain $\L = \L_1 + \L_2 = (\Y_1 + \Y_2) \Z_g = \mathrm{diag}(-2,\ 2)$ has an eigenvalue at $-2$, which is exactly inside the forbidden region depicted in Fig.~\ref{fig:combined_forbidden_regions}, meaning that combinations of different specifications can't be used at a single frequency.